\documentclass[twocolumn,showpacs,showkeys,amsmath,amssymb,superscriptaddress,floatfix,nofootinbib]{revtex4}

\usepackage{graphicx}
\usepackage{bm} 
\usepackage{subfigure}

\makeatletter
\newcommand\erfc{\mathop{\operator@font erfc}\nolimits}
\def\slashchar#1{\setbox0=\hbox{$#1$}
   \dimen0=\wd0 \setbox1=\hbox{/} \dimen1=\wd1
   \ifdim\dimen0>\dimen1 \rlap{\hbox to \dimen0{\hfil/\hfil}} #1
   \else  \rlap{\hbox to \dimen1{\hfil$#1$\hfil}} / \fi}

\begin{document}
 
\title{
Free-streaming approximation in early dynamics of relativistic heavy-ion collisions%
\footnote{Supported in part by the Polish Ministry of Science and Higher Education, grants N202~153~32/4247, N202~034~32/0918, and N202~249235, and by the U.S. NSF Grant No.~PHY-0653432.}}

\author{Wojciech Broniowski} 
\email{Wojciech.Broniowski@ifj.edu.pl} 
\affiliation{The H. Niewodnicza\'nski Institute of Nuclear Physics, Polish Academy of Sciences, PL-31342 Krak\'ow, Poland}
\affiliation{Institute of Physics, Jan Kochanowski University, PL-25406~Kielce, Poland} 

\author{Wojciech Florkowski} 
\email{Wojciech.Florkowski@ifj.edu.pl}
\affiliation{The H. Niewodnicza\'nski Institute of Nuclear Physics, Polish Academy of Sciences, PL-31342 Krak\'ow, Poland}
\affiliation{Institute of Physics, Jan Kochanowski University, PL-25406~Kielce, Poland} 

\author{Mikolaj Chojnacki}
\email{Mikolaj.Chojnacki@ifj.edu.pl}
\affiliation{The H. Niewodnicza\'nski Institute of Nuclear Physics, Polish Academy of Sciences, PL-31342 Krak\'ow, Poland}

\author{Adam Kisiel} 
\email{kisiel@if.pw.edu.pl}
\affiliation{Faculty of Physics, Warsaw University of Technology, PL-00661 Warsaw, Poland}
\affiliation{Department of Physics, Ohio State University, 
1040 Physics Research Building, 191 West Woodruff Ave., Columbus, OH 43210, USA}

\date{17 December 2008}

\begin{abstract}
We investigate an approximation to early dynamics in relativistic heavy-ion collisions, where after formation the partons are free streaming and around the proper time of 1~fm/c undergo a sudden equilibration described in terms of the Landau matching condition. We discuss physical and formal aspects of this approach. In particular, we show that initial azimuthally asymmetric transverse flow develops for non-central collisions as a consequence of the sudden equilibration. Moreover, the energy-momentum tensor from the free-streaming stage matches very smoothly to the form used in the transverse hydrodynamics, whereas matching to isotropic hydrodynamics requires a more pronounced change in the energy-momentum tensor. After the hydrodynamic phase statistical hadronization is carried out with the help of {\tt THERMINATOR}. The physical results for the transverse-momentum spectra, the elliptic-flow, and the Hanbury-Brown--Twiss correlation radii, including the ratio $R_{\rm out}/R_{\rm side}$ as well as the dependence of the radii on the azimuthal angle (azHBT), are properly described within our approach. The agreement is equally good for a purely hydrodynamic evolution started at an early proper time of 0.25~fm/c, or for the free streaming started at that time, followed by the sudden equilibration at $\tau \sim 1$~fm/c and then by perfect hydrodynamics. Thus, the inclusion of free streaming allows us to delay the start of hydrodynamics to more realistic times of the order of 1~fm/c. 
\end{abstract}



\maketitle 

%%%%%%%%%%%%%%%%%%%%%%%%%%%%%%%%%%%%%%%%%%%%%%%%%%%%%%%%%%%%%%%%%%%%%%%%%%%%%%%%%%

\section{Introduction \label{sec:intro}}

The heavy-ion data collected in the experiments at the Relativistic Heavy-Ion Collider (RHIC) suggest that the matter produced in relativistic heavy-ion collisions equilibrates very fast (presumably within a fraction of 1~fm/c) and its subsequent behavior is very well described by the dynamics of a perfect fluid \cite{Kolb:2003dz,Huovinen:2003fa,Shuryak:2004cy,Teaney:2001av,Eskola:2005ue,Hama:2005dz,Hirano:2007xd,Nonaka:2006yn}. The most common explanation of these features is the assumption that the formed matter is a strongly coupled quark-gluon plasma (sQGP) \cite{Shuryak:2004kh}. Another popular explanation assumes that the plasma is weakly interacting, however the plasma instabilities lead to a fast isotropization of matter, which in turn helps to achieve equilibration \cite{Mrowczynski:2005ki}. In this scenario one argues that hydrodynamics is applicable already at the time when the system is isotropic, but not necessarily equilibrated \cite{Arnold:2004ti}. Yet another explanation is based on the fully 3+1 dimensional parton cascade that includes inelastic pQCD-based bremsstrahlung and its back reaction \cite{Xu:2004mz,Xu:2008dv}. 

Differences in various theoretical approaches to the early-stage dynamics reflect our lack of precise knowledge concerning the mechanism of particle production and their early evolution. Moreover, our recent calculation reproducing consistently and uniformly the hadronic spectra, the elliptic flow, and the HBT correlation radii  \cite{Broniowski:2008vp} 
stresses the importance of the detailed shape of the initial condition from which hydrodynamics starts.
Hence, the uncertainty concerning the dynamics seems to be intertwined with the uncertainty concerning the initial conditions. Clearly, both the early dynamics and the initial conditions should be eventually obtained from the early microscopic dynamics, such as e.g. the Color Glass Condensate theory \cite{McLerran:1993ni,McLerran:1993ka,Kharzeev:2001yq}. In practice, however, the modeling of the partonic stage 
at the required precision is a very difficult task. 

The concept of early thermalization is especially intriguing. In some cases, in order to obtain a consistent description of the particle spectra and femtoscopy, the hydrodynamic model requires the initialization time as short as 0.1~fm/c \cite{Pratt:2008qv}. The commonly used argument for the early thermalization phenomenon is that the spatial eccentricity, resulting from the spatial transverse asymmetry of the colliding nuclei at non-zero impact parameters and driving the formation of the elliptic flow, decreases with time, hence, the experimentally observed large values of $v_2$ require an early onset of hydrodynamics and, consequently, an equilibrated state.  
In this paper we carefully reexamine this point of view. We investigate in detail an approximation to the early-stage dynamics in relativistic heavy ion collisions consisting of the free-streaming (FS) of partons followed by a sudden equilibration (SE) to a thermalized phase, which subsequently undergoes a hydrodynamic evolution. This FS+SE approximation has been proposed by Kolb, Sollfrank, and Heinz \cite{Kolb:2000sd}. It has been further considered in an investigation of the isotropization problem by Jas and Mrowczynski \cite{Jas:2007rw}, as well as elaborated in the context of the early development of flow by Sinyukov, Gyulassy, Karpenko, and Nazarenko \cite{Sinyukov:2006dw,Gyulassy:2007zz,Sinyukov:qm08}. 

The FS+SE approach assumes that after the formation stage the partons are first free streaming and later, around the proper time of 1~fm/c, undergo a sudden equilibration described in terms of the Landau matching condition. We discuss the physical and formal aspects of this approach, which is the basic goal of this work. In particular, we show that for non-central collisions, where the system develops spatial azimuthal anisotropy, an initial azimuthally asymmetric transverse flow develops as a consequence of FS+SE. Moreover, we show that the energy-momentum tensor obtained from the free-streaming stage matches very smoothly to the form needed for the transverse hydrodynamics, where the longitudinal pressure vanishes \cite{Bialas:2007gn}. The inclusion of the partonic free streaming starting at the proper time of about 0.25~fm/c followed by the sudden equilibration allows us to delay the start of hydrodynamics to comfortable times of the order of 1~fm/c. In the calcultations presented in this paper we use the isotropic perfect hydrodynamics, as described in Ref.~\cite{Broniowski:2008vp}. After the hydrodynamic phase the statistical hadronization \cite{Broniowski:2001we,Baran:2003nm} is carried out with the help of {\tt THERMINATOR} \cite{Kisiel:2005hn}. The obtained physical results for the transverse-momentum spectra, the elliptic-flow, and the Hanbury-Brown--Twiss correlation radii, including the ratio $R_{\rm out}/R_{\rm side}$ as well as the dependence of the radii on the azimuthal angle (azHBT) \cite{Kisiel:2008ws}, are all properly described within our approach. Thus, the approach consisting of FS+SE followed by hydrodynamics from $\tau=1$~fm/c may be used to obtain the uniform description of the soft hadronic data in a very similar way as in Ref. \cite{Broniowski:2008vp}, where the hydrodynamic evolution starts right away 
at the early proper time of $\tau_0 = 0.25$~fm/c and no free streaming is present. 

The outline of the paper is as follows: In the next Section we introduce the basic concepts, the physical interpretation, and the 
kinematics of the FS+SE approach. In Sect. III the structure of the energy-momentum tensor of free-streaming particles is analyzed in detail to show how the asymmetric flow is generated in this framework. The Landau matching condition is worked out in Sect. IV, while the physical results obtained with the hydrodynamic and statistical-hadronization codes are presented in Sect. V. We conclude in Sect. VI. Throughout the paper we use the units where $c=\hbar=1$. 

\section{Free streaming followed by sudden equilibration}

\begin{figure}[tb]
\subfigure{\includegraphics[angle=0,width= .48\textwidth]{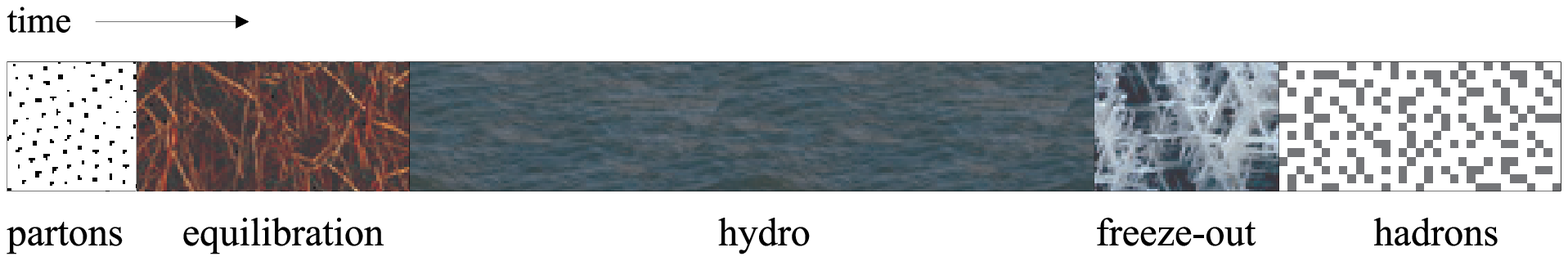}}\\
\subfigure{\includegraphics[angle=0,width= .48\textwidth]{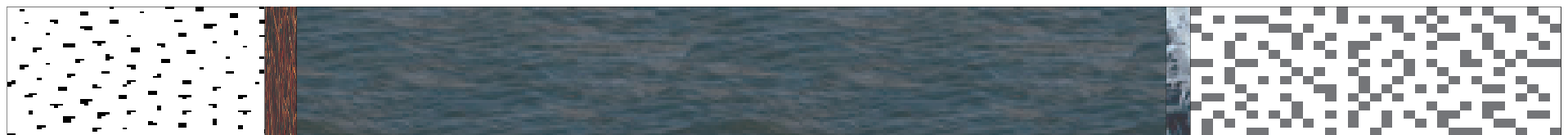}}
\caption{(Color online) Schematic history of a system formed at mid-rapidity in relativistic heavy-ion collisions, consisting of partonic free streaming, equilibration, hydrodynamics, freeze-out, and free streaming of hadrons to detectors. In the top panel the equilibration as well as freeze-out occur gradually, while in the approximate scheme depicted in the lower panel they occur suddenly. \label{fig:history}}
\end{figure}

\subsection{Basic idea} 

The basic idea to idealize the early stage of evolution of a system formed in heavy-ion collisions by a stage of collisionless partonic free-streaming (FS) followed by sudden equilibration (SE), and then by hydrodynamics, has been proposed by Kolb, Sollfrank, and Heinz \cite{Kolb:2000sd} several years ago in the context of the development of azimuthally asymmetric flow. We visualize this approximation in Fig. \ref{fig:history}. The approach assumes a sudden but delayed transition from a non-equilibrium initial state, consisting
of free-streaming partons, to a fully thermalized fluid. Ever since it has been generally thought that the approach, which admittedly decreases 
the spatial asymmetry with time, leads automatically to a reduction of the elliptic flow, which hydrodymically develops from the azimuthal asymmetry of the density profile. However, the mechanism is subtle. While free streaming {\em itself} cannot generate azimuthal asymmetry in the momentum distribution, according to the common knowledge that interactions among produced particles are needed to achieve this goal, the sudden equilibration preceded by FS is in fact capable of developing azimuthally asymmetric flow. The point is that SE is a dynamical act, where the energy-momentum tensor of the system changes abruptly into a diagonal form (in the reference frame co-moving with the fluid element). That way space-flow velocity correlations are induced, which results in a collective elliptic flow, further enhanced by the subsequent hydrodynamic evolution. 
We discuss this crucial issue in a greater 
detail in the proceeding Sections, where we are equipped with the necessary formalism, in particular in Sect.~\ref{sec:match}.

\begin{figure}[tb]
\subfigure{\includegraphics[angle=0,width= .16\textwidth]{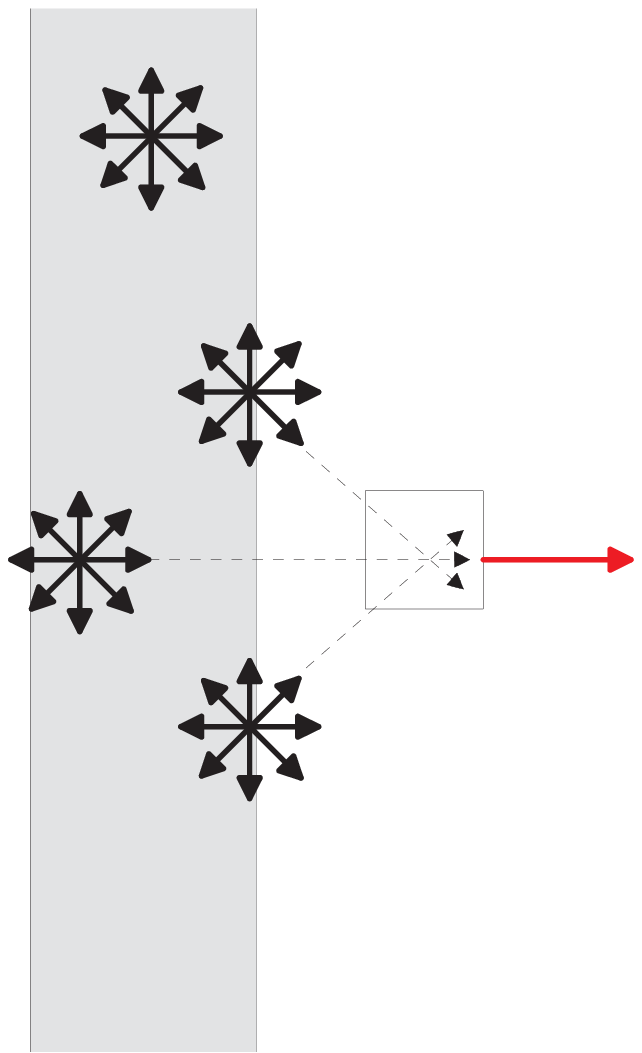}} ~~~~~~
\subfigure{\includegraphics[angle=0,width= .19\textwidth]{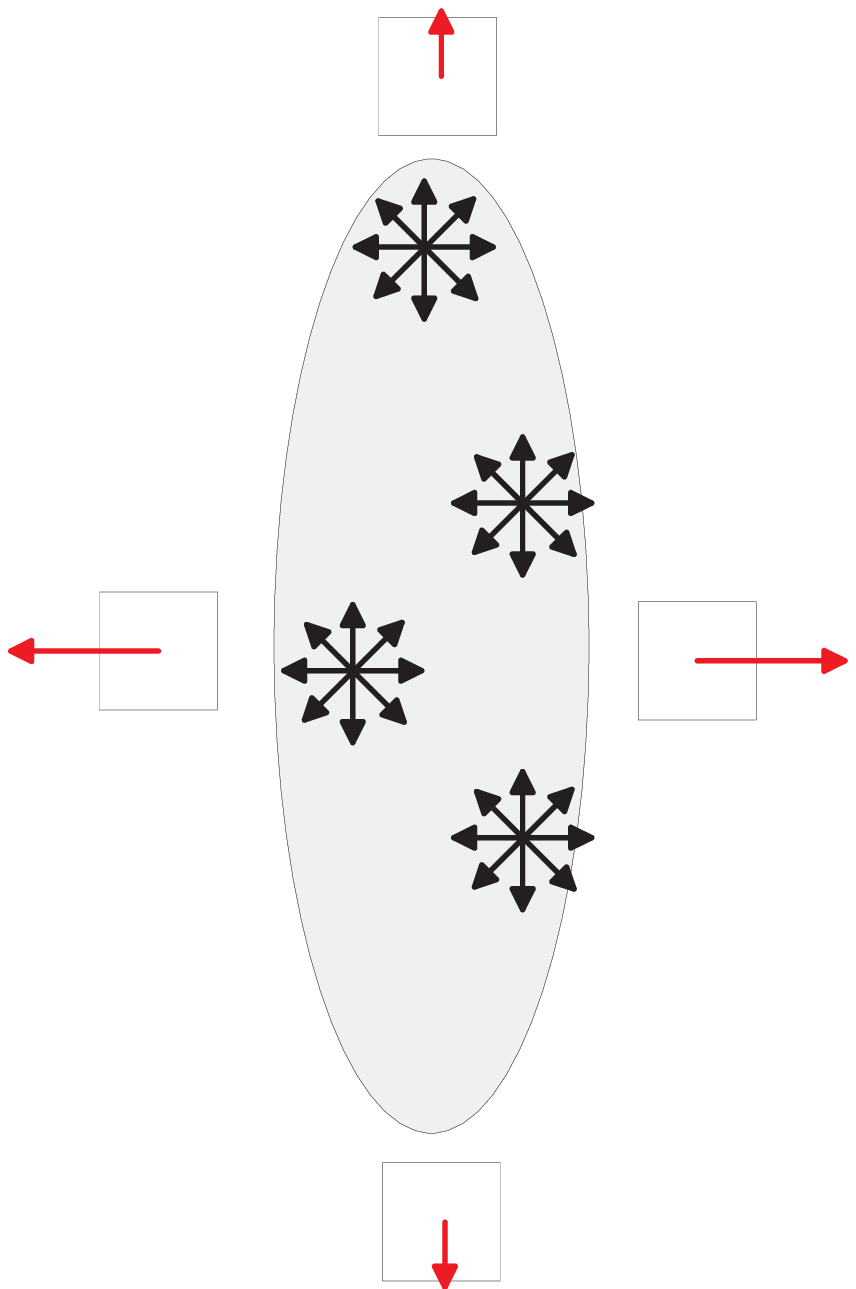}}
\caption{(Color online) Schematic view of the development of azimuthally asymmetric flow from azimuthally asymmetric systems 
 in the FS+SE approach. Arrows next to volume elements indicate the collective flow velocity. \label{fig:slab}}
\end{figure}

At this point we only wish to 
provide a qualitative argument for the development of flow in the FS+SE approach. Consider an infinite slab shown in the left part of Fig.~\ref{fig:slab}, which emits  particles isotropically  from sources denoted by dots. After some time particles reach the volume element indicated by a box. For symmetry reasons their average velocity, indicated by arrows, is perpendicular to the surface of the slab. Now, matching to perfect hydrodynamics means that we are going to treat the fluid element collectively. In other words, one may imagine that the particles glue together and move collectively in the direction perpendicular to the slab. That way a correlation of position and flow velocity is generated. In a more realistic situation of the right-hand side of Fig.~\ref{fig:slab} one starts from an ellipsoidal distribution of sources. In that case one gets a larger flow along the direction of the shorter half-axis. A quantitative calculation is presented in the following sections. The point we wish to make here is the geometric origin of the azimuthally asymmetric flow in the FS+SE approximation, which reflects the original spatial asymmetry. As a result, for non-central collisions the starting condition of hydrodynamics, when delayed with FS+SE, includes the azimuthally asymmetric initial flow velocity.

We remark that the inclusion of the non-zero transverse flow at the starting point of hydrodynamics is one of the possible ways to solve the RHIC HBT puzzle. This idea was first put forward in Ref. \cite{Chojnacki:2004ec} in the context of the thermal (hydro-inspired) models. Then it was discussed in Refs.~\cite{Gyulassy:2007zz,Sinyukov:qm08}. Quite recently, the importance of the initial flow has been strongly emphasized in Refs.~\cite{Lisa:2008gf,Pratt:2008qv}.

\subsection{Physical interpretation}

The simplest interpretation of the FS+SE approach is to simply view it as an approximation to viscous hydrodynamics. Indeed, instead of considering a complicated viscous system far from equilibrium, where microscopically the scattering cross section of partons  has a finite value, one employs an idealization, where initially the partons are free, and later develop an infinite cross section, which results in a sudden equilibration of the system. Certainly, the approach may work when viscosity decreases with time, or equivalently, the cross section increases. At first, it may seem quite paradoxical that a system which with time gets more dilute, becomes more and more likely to interact. However, 
recall that as the system gets more dilute, the average distance between partons grows, and as a result the strong coupling constant between colored objects increases. These confinement effects would make the partons in the system more likely to interact as the time goes on. Admittedly, as mentioned in the Introduction, the issue of thermalization is rather complicated, as it is difficult to assess if the initial gluon system, which is born very far from equilibrium, has enough time to equilibrate before falling apart due to expansion.

\subsection{Kinematics}

\begin{figure}[tb]
\begin{center}
\includegraphics[angle=0,width=0.36 \textwidth]{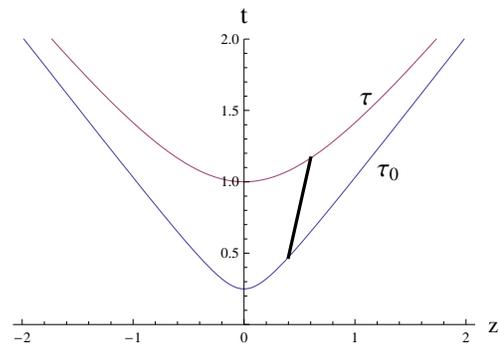}
\end{center}
\caption{(Color online) Straight-line trajectory of the parton from the proper time $\tau_0$ to the SE time $\tau$.  
\label{fig:pltau}  }
\end{figure}

Massless partons are formed at the initial proper time  $\tau_0=\sqrt{t_0^2-z_0^2}$ and move along straight lines at the speed of light until the proper time when free streaming ends, $\tau=\sqrt{t^2-z^2}$ (cf. Fig~\ref{fig:pltau}). We introduce the space-time rapidities \mbox{$\eta_0={1\over2}\log{{t_0-z_0}\over{t_0+z_0}}$} and \mbox{$\eta={1\over2}\log{{t-z}\over{t+z}}$}. Elementary kinematics, following simply from the fact that the particles move along straight lines with the velocity of light, links the positions of a parton on the initial and final hypersurfaces and its four-momentum \mbox{$p^\mu=(p_T {\rm cosh} Y, p_T \cos \phi, p_T \sin \phi, p_T {\rm sinh} Y)$}, where $Y$ and $p_T$ are the parton's rapidity and transverse momentum. We find
\begin{eqnarray}
&&\tau {\rm sinh}(\eta-Y)=\tau_0 {\rm sinh}(\eta_0-Y),  \label{kinem} \\
&&x=x_0+d \cos \phi, \;\;  y=y_0 + d\sin \phi  , \nonumber \\
&& d =\frac{t-t_0}{{\rm cosh}Y}=\tau {\rm cosh}(Y-\eta)-\sqrt{\tau_0^2+\tau^2 {\rm sinh}^2(Y-\eta)}. \nonumber
\end{eqnarray} 
The same equations are derived in \cite{Sinyukov:2006dw} through the use of the collisionless Boltzmann equation. Due to Eqs.~(\ref{kinem}) the phase-space densities of partons at the proper times $\tau_0$ and $\tau$ are related,
\begin{eqnarray}
&&\frac{d^6N(\tau)}{dY d^2p_T d\eta dx dy} = \int d \eta_0 dx_0 dy_0 \frac{d^6N(\tau_0)}{dY d^2p_T d\eta_0 dx_0 dy_0} \times \nonumber \\ 
&& \delta(\eta_0-Y-{\rm arcsinh} [\frac{\tau}{\tau_0} {\rm sinh}(\eta-Y)] ) \times \label{fs} \\
&&\delta(x - x_0 - d  \cos \phi)\delta( y - y_0 - d \sin \phi). \nonumber
\end{eqnarray}  

\subsection{Narrowing of the rapidity distribution with time}

\begin{figure}[tb]
\subfigure{\includegraphics[angle=0,width= .45\textwidth]{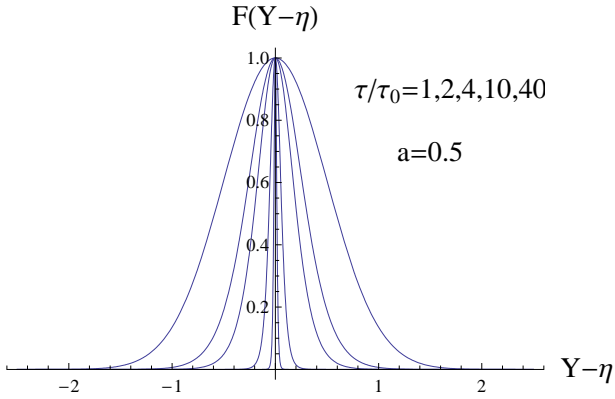}}
\caption{(Color online) The sharpening with time of the rapidity distribution of partons emitted at the space-time rapidity $\eta_0$, resulting from the free-streaming kinematics. The plotted functions are normalized to unity at the origin. At large proper times $\tau$ 
the distribution approaches $\delta(Y-\eta)$. \label{fig:narrow}}
\end{figure}

It is reasonable to assume for simplicity a factorized boost-invariant form of the initial distribution of partons, 
\begin{eqnarray}
 \frac{d^6N(\tau_0)}{dY d^2p_T d\eta_0 dx_0 dy_0} = n(x_0,y_0) F(Y-\eta_0,p_T),
\end{eqnarray}
where $n$ is their density 
in the transverse plane. Below we will apply the profile obtained from the Glauber model as given by {\tt GLISSANDO} \cite{Broniowski:2007nz}, as well as a simple Gaussian profile
\begin{eqnarray}
n(x_0,y_0)=\exp \left ( -\frac{x_0^2}{2a^2} -\frac{y_0^2}{2 b^2} \right ), 
\label{profile}
\end{eqnarray} 
where $a$ and $b$ depend on the centrality (impact parameter) of the collision.

When the emission profile $F$ is focused near $Y=\eta_0$, for instance 
\begin{eqnarray}
F \sim \exp[-(Y-\eta_0)^2/(2 a^2)], 
\end{eqnarray}
with the rapidity width parameter $a \sim 1$, then  the kinematic condition (\ref{kinem}) 
transforms it into a function of $Y-\eta$,
\begin{eqnarray}
F \sim \exp \left ( -{{\rm arcsinh}^2 \left [ \frac{\tau}{\tau_0} \sin(Y-\eta) \right ]}/({2 a^2}) \right ). \label{dist:F}
\end{eqnarray}
As the ratio $\tau/\tau_0$ increases, the distribution (\ref{dist:F}) becomes more and more peaked, as shown in Fig.~\ref{fig:narrow}. At large values of $\tau/\tau_0$ the rapidity distribution is so sharply peaked around $Y=\eta$ that effectively 
\begin{equation}
F \sim \delta(Y-\eta). \label{delta}
\end{equation}
We note that this form is frequently assumed right away as an initial condition for subsequent evolution of the system. Here it effectively follows from the kinematics of free streaming and becomes better and better as $\tau$ increases.  
 
With the form (\ref{delta}) Eq.~(\ref{fs}) yields
\begin{eqnarray}
\frac{d^6N(\tau)}{dY d^2p_T d\eta dx dy} &=& n(x-\Delta \tau \cos \phi,y-\Delta \tau \sin \phi) \times \nonumber \\
&&\delta(Y-\eta)f(p_T). \label{e:delta}
\end{eqnarray}
where $\Delta \tau=\tau -\tau_0$ and $f(p_T)$ is a transverse momentum distribution of partons (note that $d=\Delta\tau$ for $Y=\eta$). In the calculations presented in the following sections $\tau/\tau_0=4$. We read off from Fig.~\ref{fig:narrow} that with this ratio the spread in rapidity is a fraction of unity, hence very narrow and the approximation (\ref{delta}) is well justified. 

\section{Development of asymmetric flow}

\subsection{Energy-momentum tensor from free streaming}

The energy-momentum tensor at the proper time $\tau$, rapidity $\eta$, and transverse position $(x,y)$ is given by the formula
\begin{eqnarray}
&&T^{\mu \nu}=\int dY d^2p_T  \frac{  d^6N(\tau)}{dY d^2p_T d\eta dx dy} p^\mu p^\nu  \label{tmunu} \\
&&=A \int_0^{2 \pi} d\phi \, n\left(x-\Delta \tau \cos \phi,y-\Delta \tau \sin \phi\right) \times \nonumber \\
&&\left ( \begin{array}{cccc} 
{\rm cosh}^2 \eta &  {\rm cosh}\eta \cos \phi &  {\rm cosh}\eta \sin \phi & {\rm cosh}\eta {\rm sinh}\eta\\ 
{\rm cosh}\eta \cos \phi & \cos^2 \phi & \cos \phi \sin \phi & \cos \phi {\rm sinh} \eta \\ 
{\rm cosh} \eta \sin \phi & \cos \phi \sin \phi & \sin^2 \phi & \sin \phi {\rm sinh} \eta \\
{\rm cosh}\eta {\rm sinh}\eta & \cos \phi {\rm sinh}\eta & \sin \phi {\rm sinh}\eta & {\rm sinh}^2\eta \end{array} \right ), \nonumber 
\end{eqnarray}
where $A$ is a constant coming from the $p_T$ integration, which factorizes out as a consequence of the approximations
adopted earlier. Due to the assumed boost invariance the further calculations may be carried out at $\eta=0$, where we may drop the fourth row and column containing zeros, and write
\begin{eqnarray}
&&T^{\mu \nu}=A \int_0^{2 \pi} d\phi \, n\left(x-\Delta \tau \cos \phi,y-\Delta \tau \sin \phi\right) \times \nonumber \\
&&\left ( \begin{array}{ccc} 
1         &   \cos \phi         &  \sin \phi \\ 
\cos \phi & \cos^2 \phi         & \cos \phi \sin \phi \\ 
\sin \phi & \cos \phi \sin \phi & \sin^2 \phi  \end{array} \right ). \label{eq:T0}
\end{eqnarray}

\subsection{Local rest frame \label{sec:RF}}

\begin{figure}[tb]
\begin{center}
\subfigure{\includegraphics[angle=0,width=0.42 \textwidth]{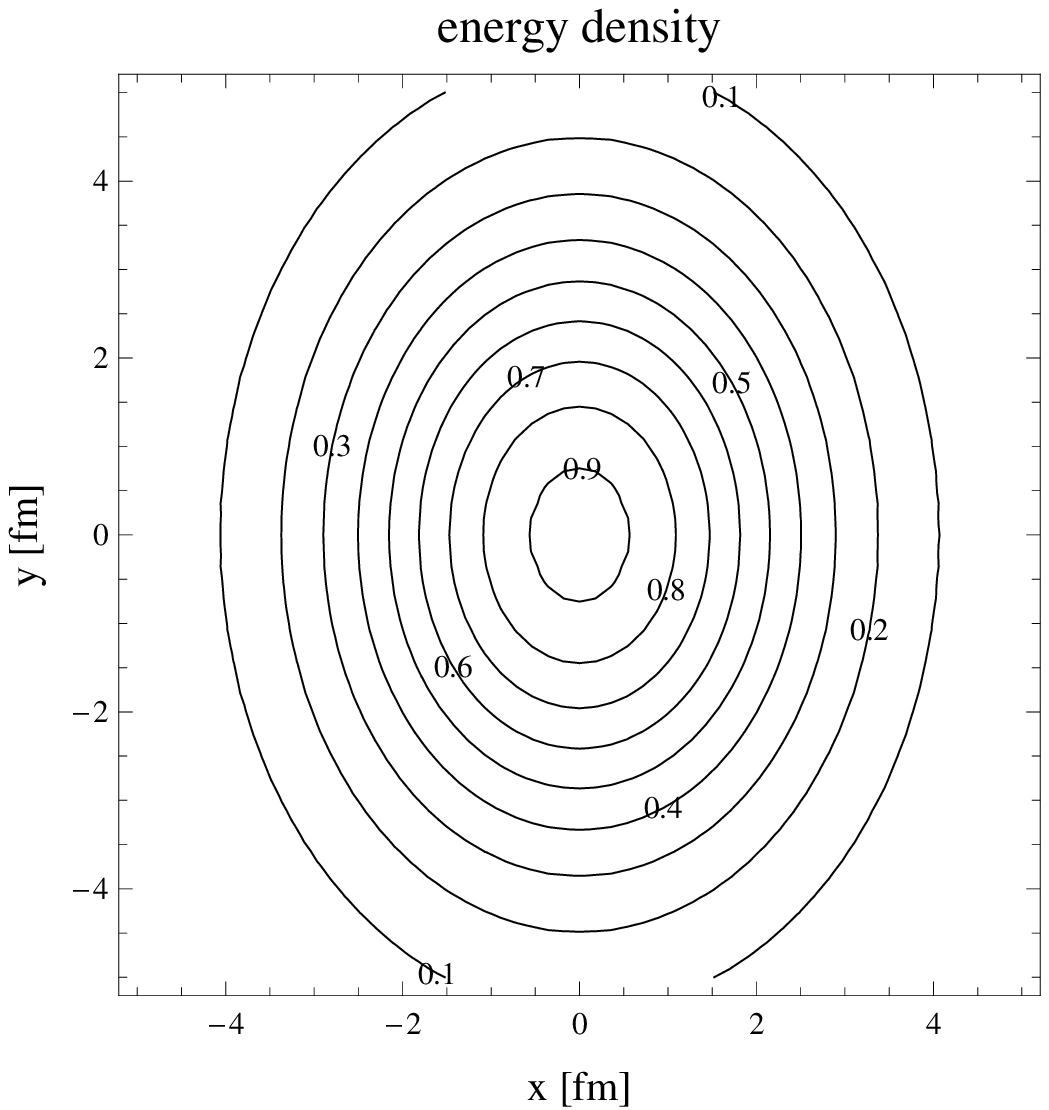}}\\
\subfigure{\includegraphics[angle=0,width=0.42 \textwidth]{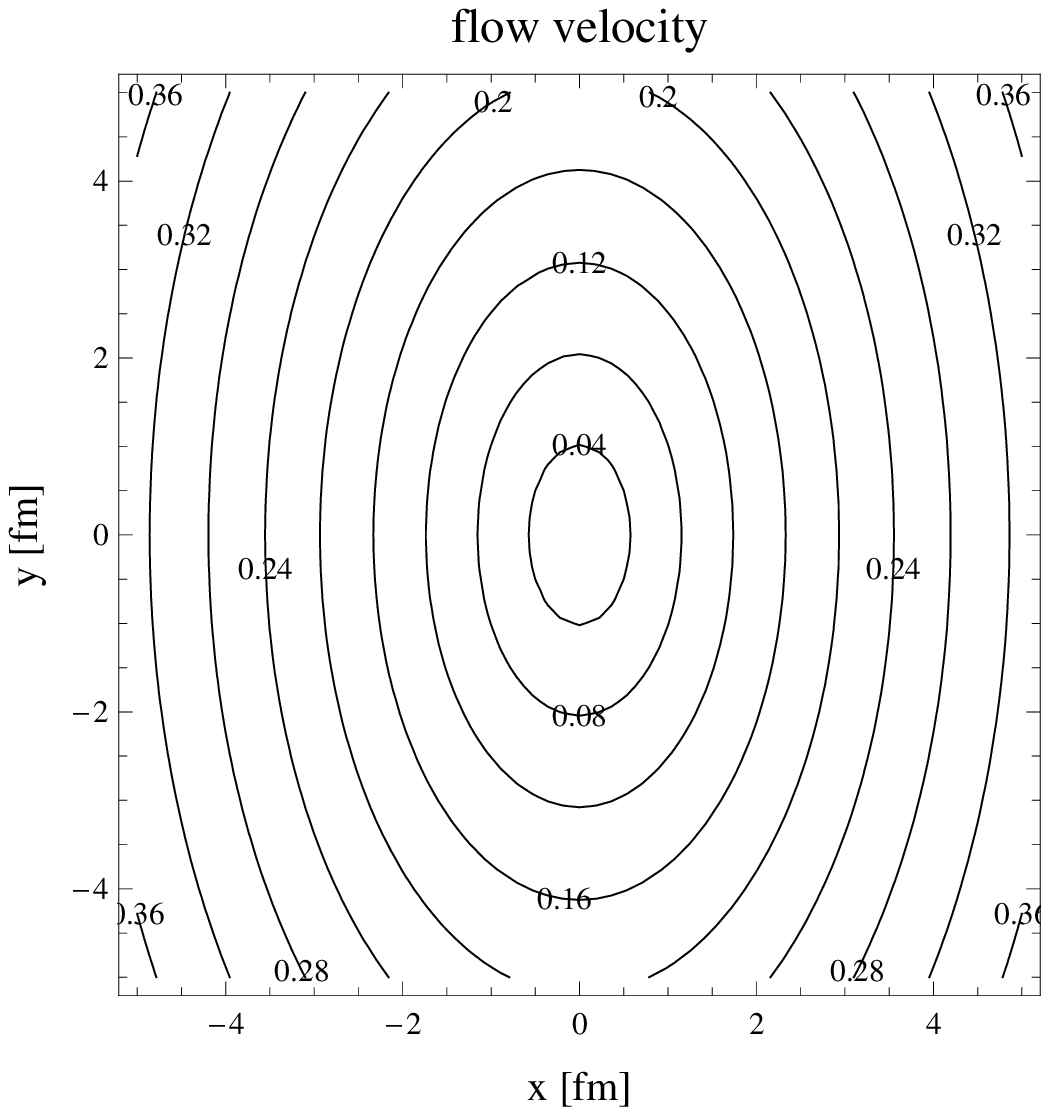}}
\end{center}
\vspace{-6.5mm}
\caption{(Color online) Contour maps of the energy-density profile $\varepsilon$ at the end of the free-streaming evolution, normalized to unity at the origin (top), and of the transverse velocity  $v=\sqrt{v^2_x+v^2_y}$ (bottom). The initial profile is from Eq.~(\ref{profile}) with $a=1.9$~fm and $b=2.5$~fm, corresponding to centrality 20-30\% for the Au+Au collisions at RHIC. The initial and final proper times of free streaming are
$\tau_0=0.25$~fm and $\tau=1$~fm. 
\label{fig:cont}}
\end{figure}

Next, at each point $(x,y)$ we pass to a local reference frame where the averaged three-momentum contained by particles in a volume element vanishes. The four-velocity of boost needed for the passage to this local rest frame is obtained from the condition  
\begin{eqnarray}
T^{\mu \nu}(x,y) u_\nu(x,y) =\varepsilon(x,y) g^{\mu \nu} u_\nu(x,y), \label{landau}
\end{eqnarray}
with
\begin{eqnarray}
u^\mu=(1,v_x,v_y,0)/\sqrt{1-v^2}, \;\;\;v=\sqrt{v_x^2+v_y^2}. \label{u}
\end{eqnarray}
Indeed, from the Lorentz covariance in the local rest frame $u_{\rm RF}^\nu(x,y)=(1,0,0)$ and Eq.~(\ref{landau}) takes the form 
\begin{eqnarray}
T_{\rm RF}^{\mu 0}(x,y) =\varepsilon(x,y) g^{\mu 0}, \label{landau2}
\end{eqnarray}
thus 
\begin{eqnarray}
T_{\rm RF}^{00}(x,y) =\varepsilon(x,y), \;\; T_{\rm RF}^{0i}(x,y) = 0, \;\;\;(i=1,2,3).\label{landau3}
\end{eqnarray}
Thus $\varepsilon$ is simply the energy density of the system in the local rest frame. 

\subsection{Energy-density and expansion velocity profiles}

We plot sample profiles of $\varepsilon$ and $v$ for a non-central collision in Fig.~\ref{fig:cont}. The plot corresponds to the proper time of $\tau=1$~fm, with the starting proper time of free streaming at $\tau_0=0.25$~fm from the Gaussian profile of Eq.~(\ref{profile}). The width parameters are \mbox{$a=1.9$~fm} and \mbox{$b=2.5$~fm}, which corresponds to centrality 20-30\% for the Au+Au collisions at the highest RHIC energy. We note that both profiles are elongated along the $y$-axes. For the energy-density it reflects the shape of the initial density profile. For the velocity we also find a steeper rise along the $x$-axis than the $y$-axis, with clear anisotropy, or the space-velocity correlation, visible.

\begin{figure}[t]
\begin{center}
\subfigure{\includegraphics[angle=0,width=0.4 \textwidth]{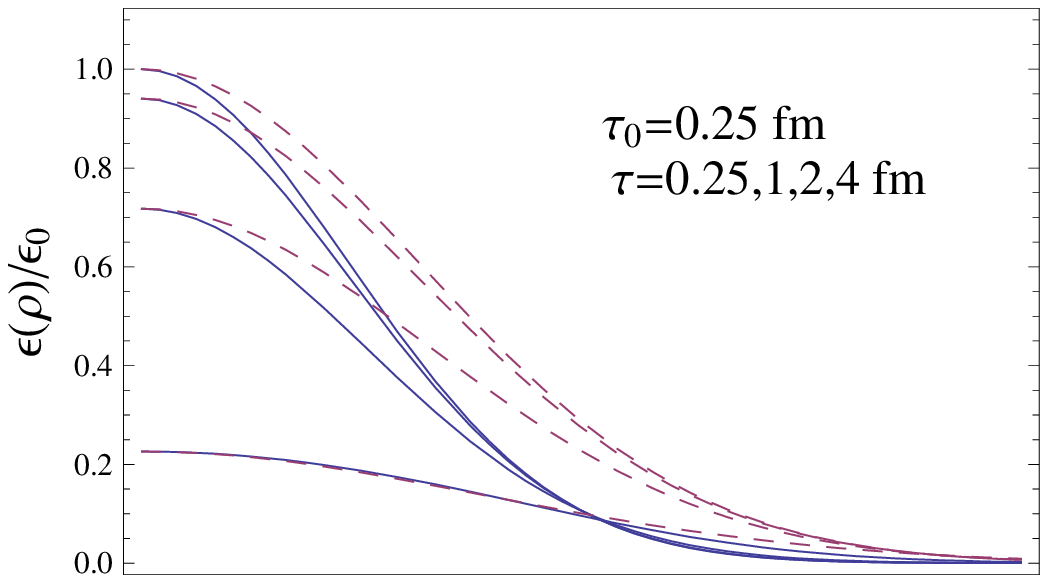}}\\
\vspace{-7mm}
\subfigure{\includegraphics[angle=0,width=0.4 \textwidth]{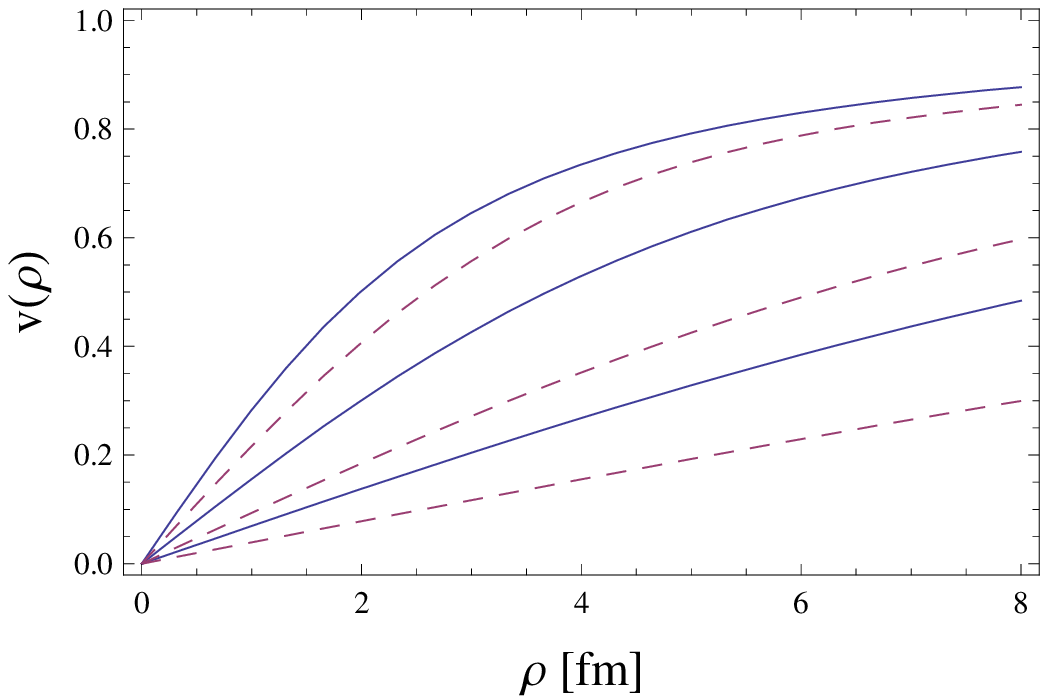}}
\end{center}
\vspace{-6.5mm}
\caption{(Color online) Sections of the energy-density profile $\varepsilon$ (top) normalized to $\varepsilon_0=\varepsilon(0,0;\tau_0)$ at the origin, and of the velocity profile $v=\sqrt{v^2_x+v^2_y}$ (bottom), cut along the $x$ axis (solid lines) and $y$-axis (dashed lines) and plotted vs. $\rho=\sqrt{x^2+y^2}$. The initial profile is from Eq.~(\ref{profile}) for centrality 20-30\% at $\tau_0=0.25$~fm. The $\varepsilon$ profiles are for $\tau=\tau_0=0.25$, $1$, $2$, and $4$~fm (from top to bottom), while the velocity profiles are for $\tau=1$, $2$, and $4$~fm (from bottom to top). We note that the flow is azimuthally asymmetric and stronger along the $x$ axis.
\label{fig:fs}}
\end{figure}

The top panel of Fig.~\ref{fig:fs} shows the sections of the energy-density profile $\varepsilon$, normalized to its value at the origin at time $\tau_0$, {\em i.e.} $\varepsilon_0=\varepsilon(0,0;\tau=\tau_0)$, plotted as functions of the transverse radius $\rho=\sqrt{x^2+y^2}$ for several values of $\tau$. The solid (dashed) lines correspond to the in-plane (out-of-plane) directions. Certainly, the system is more elongated in the out-of-plane direction. As the proper time progresses, the  system spreads out, and the value of $\varepsilon$ at the origin drops.  The corresponding velocity of flow is shown in the bottom part of Fig.~\ref{fig:fs}. We note several features: the growth of the velocity with $\tau$, the nearly linear increase with $\rho$ near the origin, and, importantly, a {\em stronger flow in the in-plane direction}. Thus the azimuthally asymmetric flow develops. This feature will be explained in Sect.~\ref{sec:diff} below.

\begin{figure}[t]
\begin{center}
\subfigure{\includegraphics[angle=0,width=0.4 \textwidth]{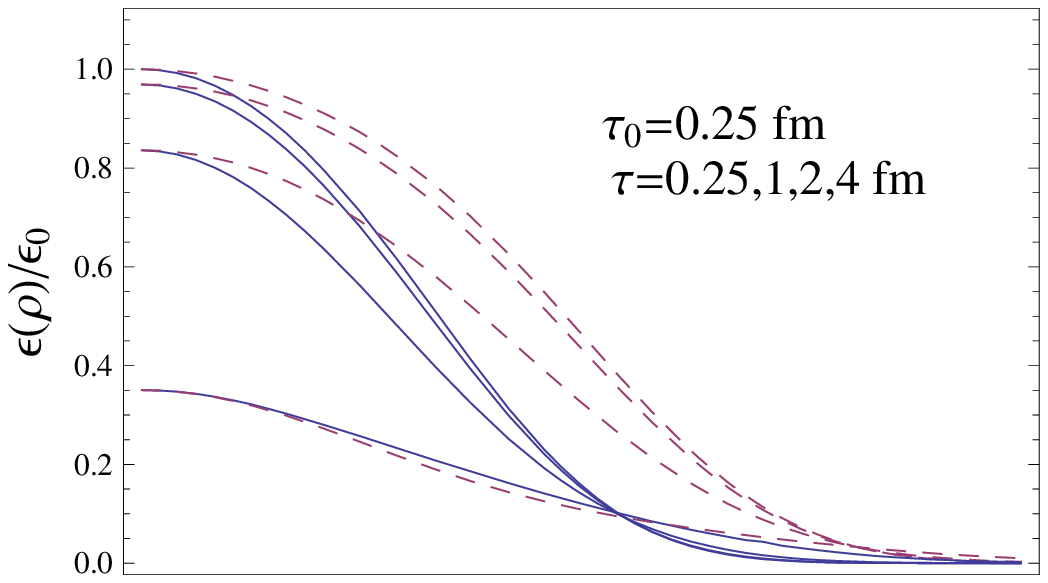}}\\
\vspace{-7mm}
\subfigure{\includegraphics[angle=0,width=0.4 \textwidth]{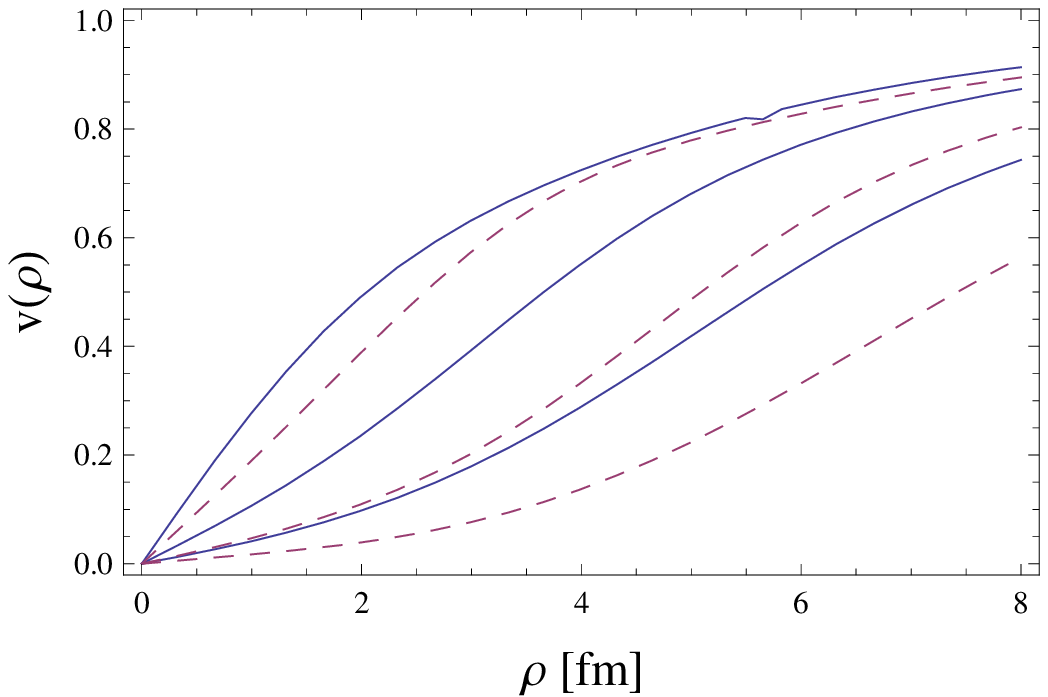}}
\end{center}
\vspace{-6.5mm}
\caption{(Color online) Same as Fig.~\ref{fig:fs} for the Glauber initial conditions from {\tt GLISSANDO}.
\label{fig:fsgl}}
\end{figure}

We have repeated the above analysis for a Glauber-model profile as obtained from the {\tt GLISSANDO} simulations \cite{Broniowski:2007nz}. The applied model is the so-called mixed model, with 85.5\% of wounded nucleons and 14.5\% of binary collisions. 
The result for the Glauber profile is shown in Fig.~\ref{fig:fsgl}. 
Comparing it to the Gaussian-profile case of Fig.~\ref{fig:fs}, 
we note qualitative similarities, but also important differences. The flow near the origin develops more slowly in the Glauber case, 
as a result the drop of $\varepsilon$ at the origin is also slower. We note that the two initial energy-density 
profiles we compare are normalized in the same way, as well as their second moments (the width parameters) are equal, {\em i.e.},
\begin{eqnarray}
&&\int dx_0 dy_0 n(x_0,y_0)={\rm const.},  \\
&&\int dx_0 dy_0 x_0^2 n(x_0,y_0)=a^2, \nonumber \\
&&\int dx_0 dy_0 y_0^2 n(x_0,y_0)=b^2. \nonumber
\end{eqnarray}
The difference shows at higher moments, with the Glauber profile being more flat at the origin than the Gaussian. As we have seen comparing Figs.~\ref{fig:fs} and \ref{fig:fsgl}, these subtle differences make an impact on the flow profile and the strength of the flow velocity. These 
issues were emphasized in Ref. \cite{Broniowski:2008vp}, where it was also shown that the obtained results, in particular $R_{\rm out}/R_{\rm side}$, 
are better when the Gaussian profile is employed.

\subsection{Small time- and gradient expansion \label{sec:diff}}

The qualitative features of the behavior presented above may be understood in terms of the low $\Delta \tau$ and low $\rho$ expansion. Expanding to lowest order in $\Delta \tau$,
\begin{eqnarray}
&& n(x-\Delta \tau \cos \phi, y-\Delta \tau \sin \phi)=  \\
&& n(x,y)-\partial_x n(x,y)\Delta \tau \cos \phi  -\partial_y n(x,y)\Delta \tau \sin \phi, \nonumber
\end{eqnarray}
and integrating over $\phi$ in Eq.~(\ref{eq:T0}), yields the energy-momentum tensor in the form
\begin{eqnarray}
T^{\mu \nu}=A \left ( \begin{array}{ccc} 
n         &   -\frac{1}{2}\Delta \tau \partial_x n         &   -\frac{1}{2}\Delta \tau \partial_y n \\ 
-\frac{1}{2}\Delta \tau \partial_x n & \frac{1}{2} n         & 0 \\ 
 -\frac{1}{2}\Delta \tau \partial_y n & 0 & \frac{1}{2} n  \end{array} \right ). \label{eq:T00}
\end{eqnarray}
The solution of Eq.~(\ref{landau}) gives to lowest order the eigenvector $u =(1,{\bf v})$, with the transverse velocity of the simple form
\begin{eqnarray}
{\bf v}(x,y)=-\frac{\Delta \tau}{3} \frac{\nabla n(x,y)}{n(x,y)}. \label{eq:vvv}
\end{eqnarray}
For the Gaussian profile (\ref{profile}) this immediately results in the Hubble flow
\begin{eqnarray}
{\bf v}(x,y)=\frac{\Delta \tau}{3} \left ( \frac{x}{a^2}, \frac{y}{b^2}\right ).
\end{eqnarray}
This behavior is clearly seen in Fig.~\ref{fig:fs} near the origin. The validity of Eq.~(\ref{eq:vvv}) requires the condition \mbox{$\Delta \tau {\mid \nabla n(x,y) \mid}/{n(x,y)} \ll 1$}.

\section{Landau matching} 

\subsection{Matching to transverse and isotropic hydrodynamics\label{sec:match}}

\begin{figure}[tb]
\begin{center}
\includegraphics[angle=0,width=0.42 \textwidth]{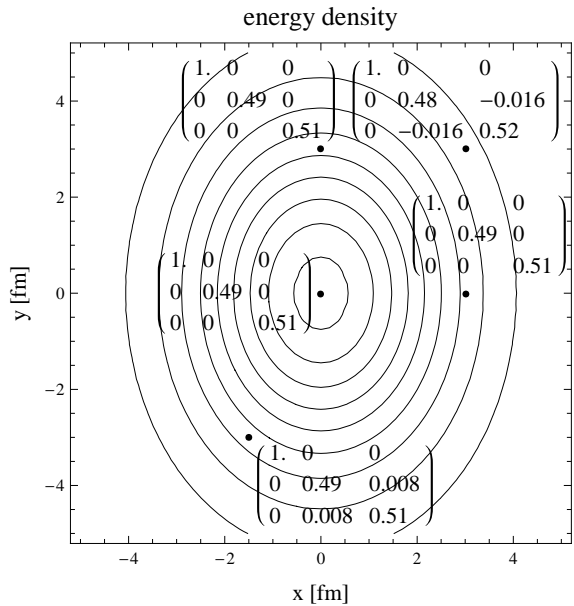}
\end{center}
\vspace{-6.5mm}
\caption{(Color online) 
\label{fig:ploT} Same as Fig.~\ref{fig:cont} with the matrix $T_{\rm RF}/\varepsilon$ shown in a few points indicated by blobs.}
\end{figure}

As already stated in Sect.~\ref{sec:RF}, the boost of $T^{\mu \nu}$ with the velocity found from Eq.~(\ref{landau}) yields $T_{RF}^{\mu \nu}$, {\em i.e.} the energy-momentum tensor in the local rest frame which satisfies the conditions (\ref{landau3}). 
In Fig.~\ref{fig:ploT} we have overlaid over the energy profile of Fig.~\ref{fig:cont} the explicit form of the matrix $T_{\rm RF}/\varepsilon$ in a few points. We note that $T_{\rm RF}$ is very close to the diagonal form 
\begin{eqnarray}
T_{\rm RF}^{\mu \nu} \simeq \varepsilon \left(
\begin{array}{lll} 1 & 0 & 0  \\
0 & \frac{1}{2} & 0 \\
0 & 0 & \frac{1}{2}
\end{array} \right) . \label{eq:T2}
\end{eqnarray}
In fact, for symmetry reasons $T_{\rm RF}$ is diagonal along the $x$ and $y$ axes, and away from them it develops only small non-diagonal pieces. Also, the difference between $T^{xx}$ and $T^{yy}$ is small, at the level of a few percent. Interestingly, Eq.~(\ref{eq:T2}) has precisely the structure of the energy-momentum tensor of the {\em perfect transverse hydrodynamics} of massless particles \cite{Bialas:2007gn}, with the transverse pressure equal to $\varepsilon/2$. Small departures from this form, present in our case, have the same structure as the shear 
tensor used to describe the viscosity effects in transverse hydrodynamics 
\cite{Bozek:2007di}. We notice larger deviation in $T_{xx}$ and $T_{yy}$ than in $T_{xy}$, the same effect as in viscous hydrodynamics.

The Landau matching condition amounts to replacing the free-streaming energy-momentum tensor to the form from perfect hydrodynamics. One may match to the perfect transverse hydrodynamics, where we replace
\begin{eqnarray}
T_{\rm RF}^{\mu \nu} \to T_2^{\mu \nu} \equiv \varepsilon \left ( \begin{array}{cccc} 1 & 0 & 0 & 0\\ 0 & 1\over 2 & 0 & 0\\ 
0 & 0 & 1\over 2 & 0 \\ 0 & 0 & 0 & 0\end{array}\right ), \label{eq:match2}
\end{eqnarray}
or to the perfect isotropic hydrodynamics, in which case  
\begin{eqnarray}
T_{\rm RF}^{\mu \nu} \to T_3^{\mu \nu} \equiv \varepsilon \left ( \begin{array}{cccc} 1 & 0 & 0 & 0\\ 0 & 1\over 3 & 0 & 0\\ 
0 & 0 & 1\over 3 & 0 \\ 0 & 0 & 0 & 1\over 3 \end{array}\right ).  \label{eq:match3}
\end{eqnarray}
Admittedly, the matching (\ref{eq:match2}) requires only a minor modification of the energy-momentum tensor, while (\ref{eq:match3}) employs a more significant change. This means that the free-streaming approximation may be rather smoothly linked to the perfect transverse hydrodynamics.

\subsection{Early generation of elliptic flow}

\begin{figure}[tb]
\begin{center}
\hspace{-10mm} \subfigure{\includegraphics[angle=0,width=0.34 \textwidth]{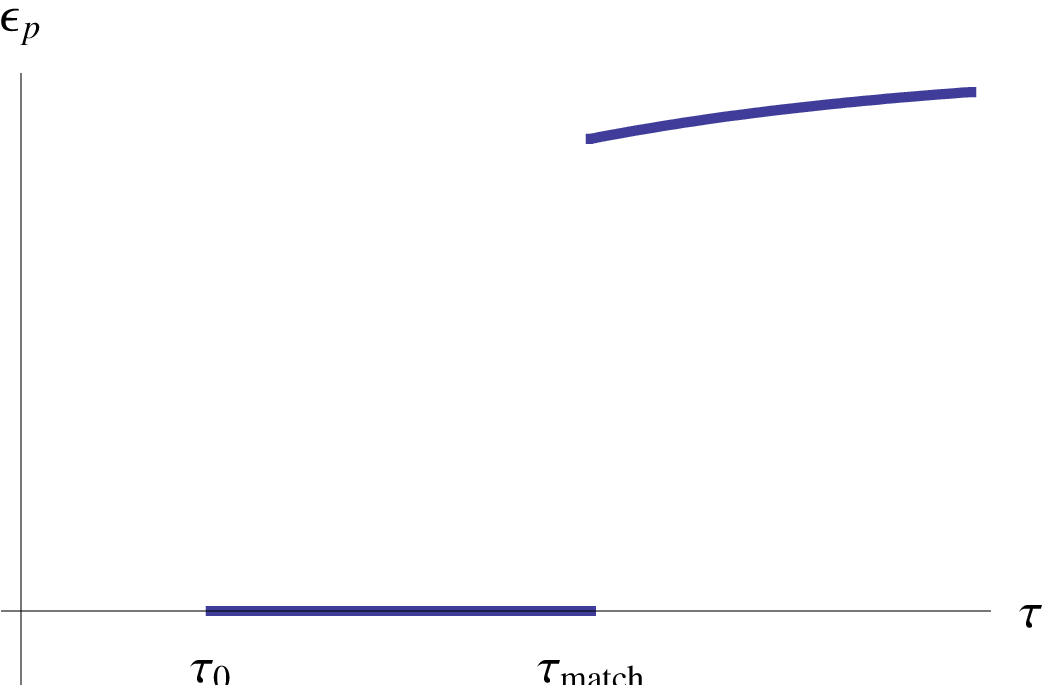}}\\
\subfigure{\includegraphics[angle=0,width=0.4 \textwidth]{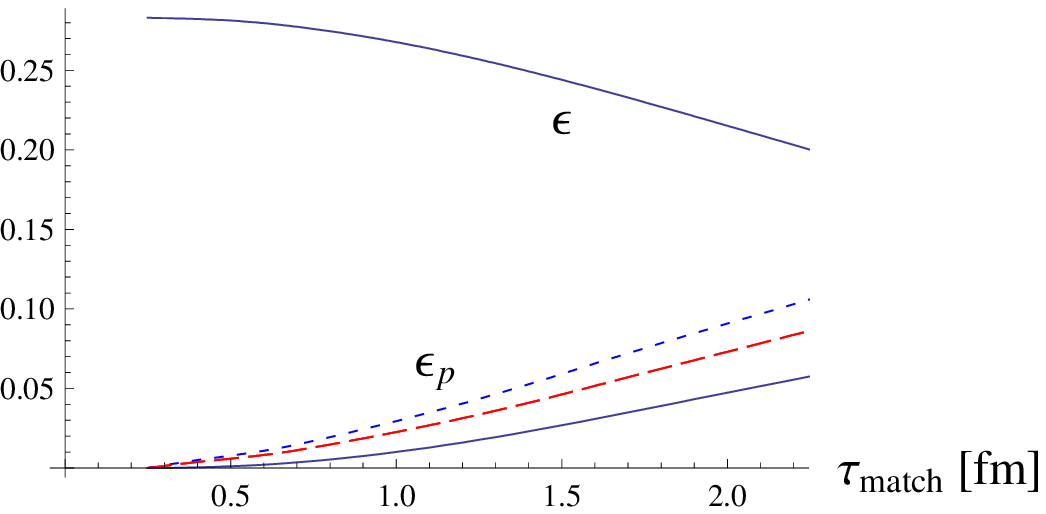}}
\end{center}
\vspace{-4.5mm}
\caption{(Color online) Top: The schematic development of the partonic elliptic flow $\epsilon_p$ from the FS+SE approximation (top). Bottom: the value of the generated momentum asymmetry $\epsilon_p$ plotted as a function of the matching proper time. 
The dotted (dashed) line corresponds to matching to isotropic (transverse) hydrodynamics, while the solid line shows 
the results of hydrodynamics only, with no free streaming. 
The top curve shows the spatial asymmetry $\epsilon$. Same parameters as in Fig.~\ref{fig:cont}.
\label{fig:match}} 
\end{figure}

We interpret the Landau matching conditions (\ref{eq:match2}) or (\ref{eq:match3}) as a {\em dynamical act}. It is not a mathematical replacement implementing an approximation. Rather, it is a simplified (recall Fig.~\ref{fig:history}) description of the interactions among partons which occur instantaneously in the FS+SE approximation. The sudden equilibration causes the development of early elliptic flow. Following Ref.~\cite{Kolb:2000sd} we consider the measure  
\begin{eqnarray}
\epsilon_p=\frac{\langle T_{xx}\rangle-\langle T_{yy}\rangle}{\langle T_{xx}\rangle+\langle T_{yy}\rangle}. \label{eq:v2}
\end{eqnarray}
The brackets indicate the space integration in the laboratory frame. Until SE occurs, $\epsilon_p = 0$, as without interactions 
the elliptic flow cannot develop, with the momentum spectrum being an uncorrelated sum over the emitting sources. At the proper time $\tau_{\rm match}$ SE interactions occur, producing as a result the energy-momentum tensor of Eq.~(\ref{eq:match2}) or (\ref{eq:match3}). This act yields immediately a non-zero $\epsilon_p$. The subsequent hydrodynamic evolution may further increase the value of the elliptic flow coefficient. The situation is depicted schematically in the top panel of Fig.~\ref{fig:match}.

In order to compute the value of $\epsilon_p$ generated by SE, we take $T_2$ or $T_3$ from Eq.~(\ref{eq:match2}) or (\ref{eq:match3}) and for each volume element we go back from its local rest frame to the laboratory frame, which is necessary in order to apply the definition (\ref{eq:v2}). The result of this procedure is shown in the bottom panel of  Fig.~\ref{fig:match}. We note that increasing the value of $\tau_{\rm match}$ results in a larger flow coefficient. Thus FS+SE {\em does generate elliptic flow}. At the same plot we also show the spatial eccentricity
\begin{eqnarray}
\epsilon=\frac{\langle y\rangle^2-\langle x\rangle^2}{\langle y\rangle^2+\langle x\rangle^2},
\end{eqnarray}
which obviously decreases with time. From the viewpoint of hydrodynamics, this decrease of spatial asymmetry is compensated by the generated asymmetric flow from FS+SE.

\section{The follow-up evolution}

\subsection{Hydrodynamics and {\tt THERMINATOR}}

The energy-momentum tensor obtained with the FS+SE approximation is plugged into the hydrodynamic
evolution. Here we use the perfect isotropic hydrodynamics, thus the matching condition (\ref{eq:match3}) is used. The equation of state uses the lattice QCD simulations of Ref.~\cite{Aoki:2005vt} at high temperatures, $T>170$~MeV, the hadronic gas at $T<170$~MeV, and a smooth interpolation in the vicinity of $170$~MeV. At the end of the hydrodynamic phase {\tt THERMINATOR} simulations are carried out in order to implement the hadronic decays. Our scheme is described fully in Refs.~\cite{Chojnacki:2007rq,Broniowski:2008vp}, so we do not provide any further details here. 

\subsection{Physical results and comparison to data \label{sec:phys}}

\begin{figure}[tb]
\begin{center}
\includegraphics[angle=0,width=0.47 \textwidth]{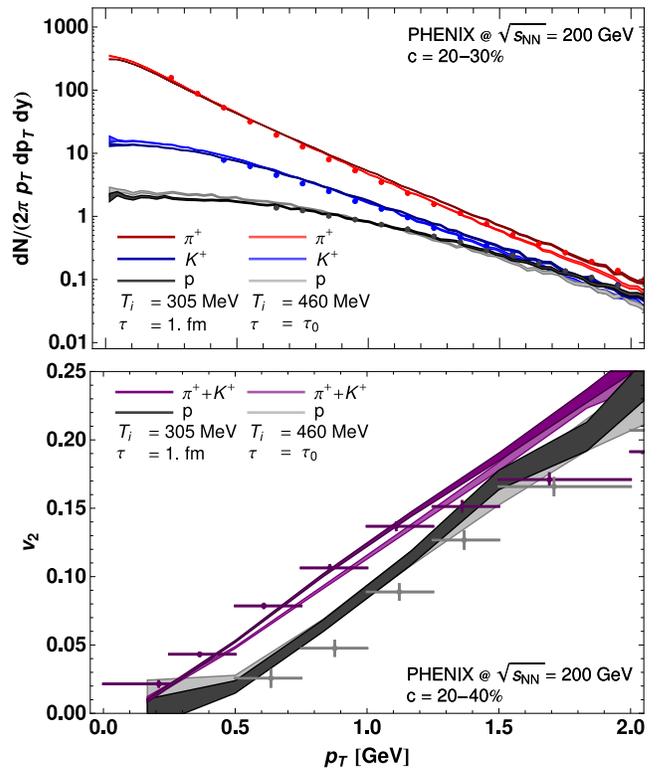}
\end{center}
\vspace{-1mm}
\caption{(Color online) The transverse-momentum spectra of pions, kaons and protons for $c$=20-30\% (upper panel) and the elliptic flow coefficient $v_2$ for  $c$=20-40\% (lower panel). The darker (lighter) lines describe the model results for the case with (without) free streaming. Data from \cite{Adler:2003cb,Adler:2003kt}.
\label{fig:spv2}}
\end{figure}

\begin{figure}[tb]
\vspace{2mm}
\begin{center}
\includegraphics[angle=0,width=0.4 \textwidth]{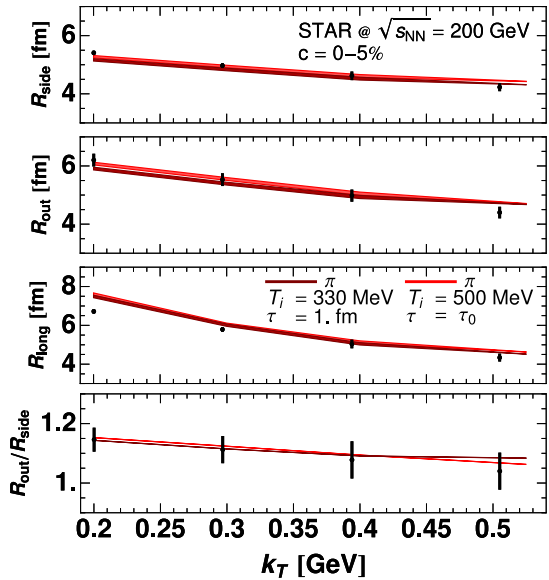}
\end{center}
\vspace{-1mm}
\caption{(Color online) The pion HBT radii $R_{\rm side}$ , $R_{\rm out}$ , $R_{\rm long}$, 
and the ratio $R_{\rm out}/R_{\rm side}$ for central collisions. The darker (lighter) 
lines describe the results with (without) FS+SE. The data from \cite{Adams:2004yc}.
\label{fig:hbt}}
\end{figure}

In the following we compare the results obtained with hydrodynamics only (starting at the proper 
time \mbox{$\tau_0=0.25$~fm}) and the results obtained with free streaming from $\tau_0$ to $\tau=1.0$~fm, followed by SE and hydrodynamics. In each case we start from the Gaussian profile (\ref{profile}). The results are shown in Figs.~\ref{fig:spv2} and \ref{fig:hbt} with darker lines indicating the calculation with FS+SE, and the lighter lines with hydrodynamics only. We notice very similar results for the two considered cases, not to mention the very good description of the data. Larger free-streaming times ($\tau-\tau_0 \sim 1.5$~fm) spoil this agreement, as the flow becomes too strong. As described in Ref.~\cite{Broniowski:2008vp}, we have achieved a uniform agreement for soft physics at RHIC. In particular, the transverse-momentum spectra, the elliptic-flow, and the HBT correlation radii, including the notorious ratio $R_{\rm out}/R_{\rm side}$, are all properly described. The azimuthally-sensitive HBT correlations \cite{Voloshin:azHBT,Lisa:2000xj} are also correctly described within our framework. In Ref. \cite{Kisiel:2008ws} we showed that our model calculations reproduce the full dependence of the HBT radii and their oscillations on the transverse-momentum and centrality.

\begin{figure}[b]
\subfigure{\includegraphics[angle=0,width= .4\textwidth]{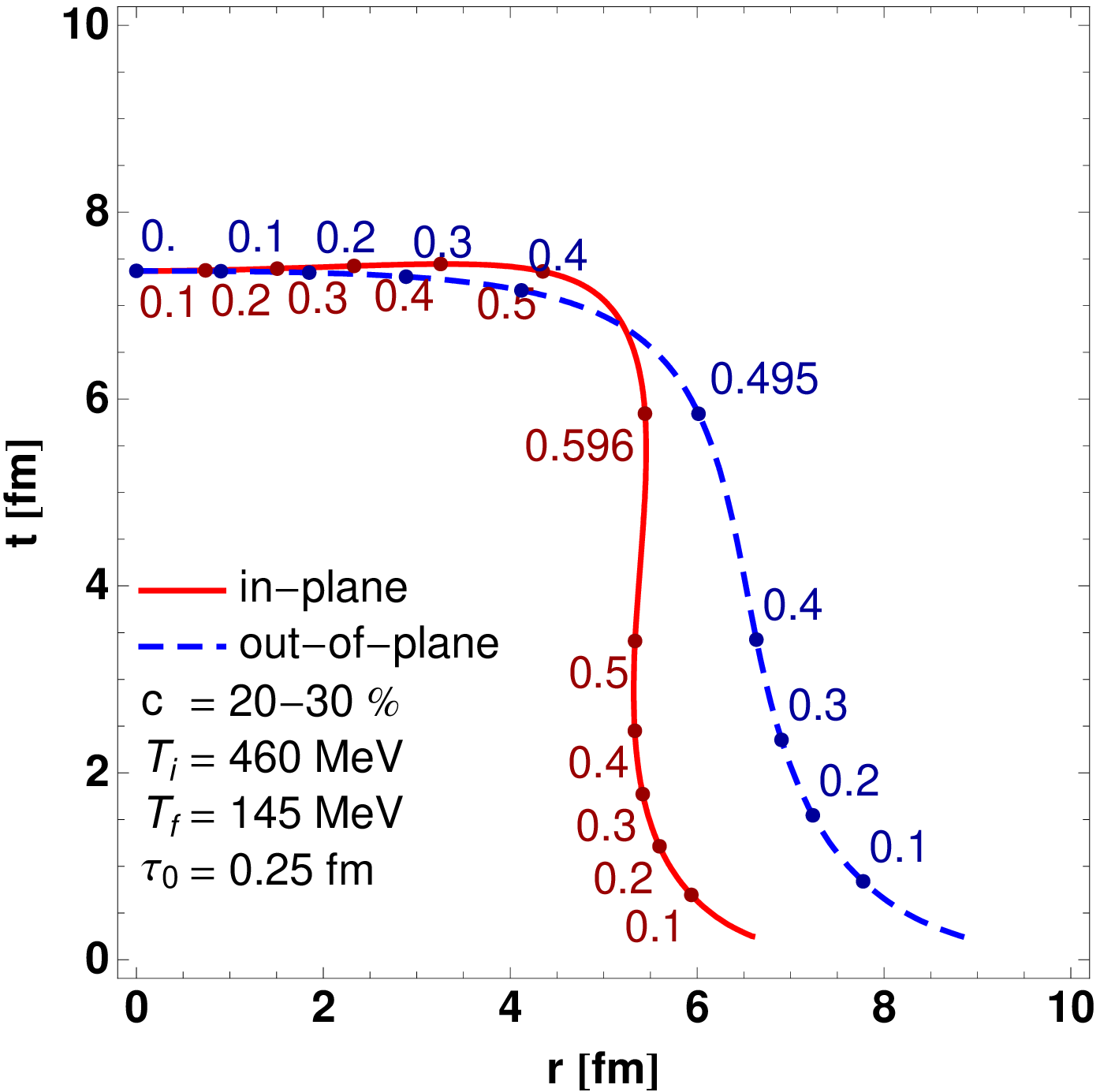}}\\
\vspace{-2mm}
\subfigure{\includegraphics[angle=0,width= .4\textwidth]{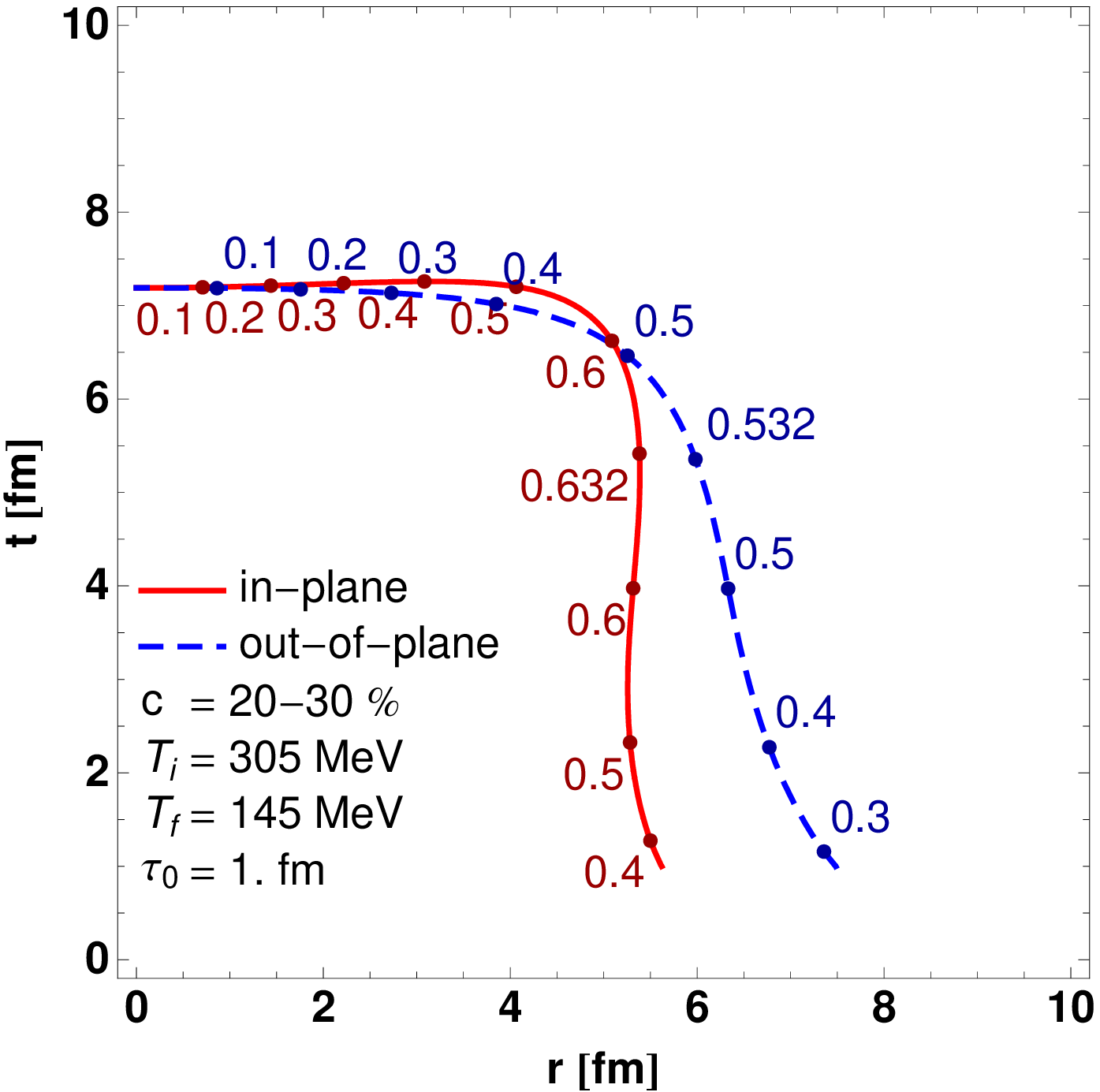}}
\caption{(Color online) The freeze-out hypersurfaces for the calculation with hydrodynamics only (top) 
and with FS+SE followed by hydrodynamics (bottom). Same parameters as in Fig.~\ref{fig:cont}. \label{fig:freeze}}
\end{figure}

It is also interesting to look at the freeze-out hypersurfaces for the two considered cases, shown in Fig.~\ref{fig:freeze}. They have very much similar shape and size, which explains again why the two schemes give essentially the same results.

The practical observation following from our study of physical observables is that the inclusion of FS+SE may be used to {\em delay the start of hydrodynamics}. The physical results are basically unaltered, since the dispersion of the density profile with time, resulting in milder hydrodynamic development of flow, is accompanied by the buildup of the {\em initial asymmetric flow}.

\section{Conclusions}

Our analysis has focused on modeling of the early stage evolution within the FS+SE approach. We stress, 
however,  that our complete model (FS + SE + hydrodynamics + statistical hadronization) describes consistently the essential features of the soft hadron production at RHIC, including the $p_T$-spectra, $v_2$, and the pionic HBT radii, including their azimuthal asymmetry. The main reasons for obtaining such a good description, listed in Ref.~\cite{Broniowski:2008vp}, were identified with the use of a realistic equation of state without the soft point, the Gaussian initial condition including the fluctuations of the initial eccentricity, as well as the inclusion of all known hadronic resonances in the statistical hadronization. This successful description of the RHIC data hints on a possible solution of the RHIC HBT puzzle~\cite{Broniowski:2008vp}. 

In this paper we have analyzed the approximation for the early-stage dynamics of relativistic heavy-ion collisions which assumes the initial free-streaming of partons followed by the sudden equilibration (FS+SE approximation). Our main findings are that such a modeling of the initial stage is compatible with the data describing soft hadron production at the highest RHIC energies. The decrease with time of the initial space asymmetry due to the free-streaming is compensated by the effects following from sudden equilibration, which lead to the formation of the radial and elliptic flow at the starting point for the hydrodynamic evolution. The net results of the FS+SE approach is that one may delay the initialization of the hydrodynamic evolution. Interestingly, unless the duration of free streaming is not larger than about 1.5~fm/c, its specific value is irrelevant for the final physical observables. Such insensitivity of the final results to the details of the initial evolution indicates that the assumption of the fast equilibration may be relaxed and replaced by a model where thermalization processes and building 
of the flow happen gradually. This issue of the persisting early thermalization puzzle might be 
resolved at a microscopic level only within an appropriate QCD-based kinetic model.

\begin{acknowledgments}
Two of us (WB and WF) are grateful to Piotr Bo\.zek and Stanis\l{}aw Mr\'owczy\'nski for useful discussions.  
\end{acknowledgments}

\bibliography{ref-rr}

\begin{thebibliography}{39}
\expandafter\ifx\csname natexlab\endcsname\relax\def\natexlab#1{#1}\fi
\expandafter\ifx\csname bibnamefont\endcsname\relax
  \def\bibnamefont#1{#1}\fi
\expandafter\ifx\csname bibfnamefont\endcsname\relax
  \def\bibfnamefont#1{#1}\fi
\expandafter\ifx\csname citenamefont\endcsname\relax
  \def\citenamefont#1{#1}\fi
\expandafter\ifx\csname url\endcsname\relax
  \def\url#1{\texttt{#1}}\fi
\expandafter\ifx\csname urlprefix\endcsname\relax\def\urlprefix{URL }\fi
\providecommand{\bibinfo}[2]{#2}
\providecommand{\eprint}[2][]{\url{#2}}

\bibitem[{\citenamefont{Kolb and Heinz}(2003)}]{Kolb:2003dz}
\bibinfo{author}{\bibfnamefont{P.~F.} \bibnamefont{Kolb}} \bibnamefont{and}
  \bibinfo{author}{\bibfnamefont{U.}~\bibnamefont{Heinz}}
  (\bibinfo{year}{2003}), \eprint{in Quark-Gluon Plasma 3, edited by R.C. Hwa
  and X.-N. Wang (World Scientific, Singapore, 2004), p. 634, nucl-th/0305084}.

\bibitem[{\citenamefont{Huovinen}(2003)}]{Huovinen:2003fa}
\bibinfo{author}{\bibfnamefont{P.}~\bibnamefont{Huovinen}}
  (\bibinfo{year}{2003}), \eprint{in Quark-Gluon Plasma 3, edited by R.C. Hwa
  and X.-N. Wang (World Scientific, Singapore, 2004), p. 600, nucl-th/0305064}.

\bibitem[{\citenamefont{Shuryak}(2005)}]{Shuryak:2004cy}
\bibinfo{author}{\bibfnamefont{E.~V.} \bibnamefont{Shuryak}},
  \bibinfo{journal}{Nucl. Phys.} \textbf{\bibinfo{volume}{A750}},
  \bibinfo{pages}{64} (\bibinfo{year}{2005}), \eprint{hep-ph/0405066}.

\bibitem[{\citenamefont{Teaney et~al.}(2001)\citenamefont{Teaney, Lauret, and
  Shuryak}}]{Teaney:2001av}
\bibinfo{author}{\bibfnamefont{D.}~\bibnamefont{Teaney}},
  \bibinfo{author}{\bibfnamefont{J.}~\bibnamefont{Lauret}}, \bibnamefont{and}
  \bibinfo{author}{\bibfnamefont{E.~V.} \bibnamefont{Shuryak}}
  (\bibinfo{year}{2001}), \eprint{nucl-th/0110037}.

\bibitem[{\citenamefont{Eskola et~al.}(2005)\citenamefont{Eskola, Honkanen,
  Niemi, Ruuskanen, and Rasanen}}]{Eskola:2005ue}
\bibinfo{author}{\bibfnamefont{K.~J.} \bibnamefont{Eskola}},
  \bibinfo{author}{\bibfnamefont{H.}~\bibnamefont{Honkanen}},
  \bibinfo{author}{\bibfnamefont{H.}~\bibnamefont{Niemi}},
  \bibinfo{author}{\bibfnamefont{P.~V.} \bibnamefont{Ruuskanen}},
  \bibnamefont{and} \bibinfo{author}{\bibfnamefont{S.~S.}
  \bibnamefont{Rasanen}}, \bibinfo{journal}{Phys. Rev.}
  \textbf{\bibinfo{volume}{C72}}, \bibinfo{pages}{044904}
  (\bibinfo{year}{2005}), \eprint{hep-ph/0506049}.

\bibitem[{\citenamefont{Hama et~al.}(2006)}]{Hama:2005dz}
\bibinfo{author}{\bibfnamefont{Y.}~\bibnamefont{Hama}} \bibnamefont{et~al.},
  \bibinfo{journal}{Nucl. Phys.} \textbf{\bibinfo{volume}{A774}},
  \bibinfo{pages}{169} (\bibinfo{year}{2006}), \eprint{hep-ph/0510096}.

\bibitem[{\citenamefont{Hirano et~al.}(2007)\citenamefont{Hirano, Heinz,
  Kharzeev, Lacey, and Nara}}]{Hirano:2007xd}
\bibinfo{author}{\bibfnamefont{T.}~\bibnamefont{Hirano}},
  \bibinfo{author}{\bibfnamefont{U.~W.} \bibnamefont{Heinz}},
  \bibinfo{author}{\bibfnamefont{D.}~\bibnamefont{Kharzeev}},
  \bibinfo{author}{\bibfnamefont{R.}~\bibnamefont{Lacey}}, \bibnamefont{and}
  \bibinfo{author}{\bibfnamefont{Y.}~\bibnamefont{Nara}}, \bibinfo{journal}{J.
  Phys.} \textbf{\bibinfo{volume}{G34}}, \bibinfo{pages}{S879}
  (\bibinfo{year}{2007}), \eprint{nucl-th/0701075}.

\bibitem[{\citenamefont{Nonaka and Bass}(2007)}]{Nonaka:2006yn}
\bibinfo{author}{\bibfnamefont{C.}~\bibnamefont{Nonaka}} \bibnamefont{and}
  \bibinfo{author}{\bibfnamefont{S.~A.} \bibnamefont{Bass}},
  \bibinfo{journal}{Phys. Rev.} \textbf{\bibinfo{volume}{C75}},
  \bibinfo{pages}{014902} (\bibinfo{year}{2007}), \eprint{nucl-th/0607018}.

\bibitem[{\citenamefont{Shuryak}(2004)}]{Shuryak:2004kh}
\bibinfo{author}{\bibfnamefont{E.}~\bibnamefont{Shuryak}}, \bibinfo{journal}{J.
  Phys.} \textbf{\bibinfo{volume}{G30}}, \bibinfo{pages}{S1221}
  (\bibinfo{year}{2004}).

\bibitem[{\citenamefont{Mrowczynski}(2006)}]{Mrowczynski:2005ki}
\bibinfo{author}{\bibfnamefont{S.}~\bibnamefont{Mrowczynski}},
  \bibinfo{journal}{Acta Phys. Polon.} \textbf{\bibinfo{volume}{B37}},
  \bibinfo{pages}{427} (\bibinfo{year}{2006}), \eprint{hep-ph/0511052}.

\bibitem[{\citenamefont{Arnold et~al.}(2005)\citenamefont{Arnold, Lenaghan,
  Moore, and Yaffe}}]{Arnold:2004ti}
\bibinfo{author}{\bibfnamefont{P.}~\bibnamefont{Arnold}},
  \bibinfo{author}{\bibfnamefont{J.}~\bibnamefont{Lenaghan}},
  \bibinfo{author}{\bibfnamefont{G.~D.} \bibnamefont{Moore}}, \bibnamefont{and}
  \bibinfo{author}{\bibfnamefont{L.~G.} \bibnamefont{Yaffe}},
  \bibinfo{journal}{Phys. Rev. Lett.} \textbf{\bibinfo{volume}{94}},
  \bibinfo{pages}{072302} (\bibinfo{year}{2005}), \eprint{nucl-th/0409068}.

\bibitem[{\citenamefont{Xu and Greiner}(2005)}]{Xu:2004mz}
\bibinfo{author}{\bibfnamefont{Z.}~\bibnamefont{Xu}} \bibnamefont{and}
  \bibinfo{author}{\bibfnamefont{C.}~\bibnamefont{Greiner}},
  \bibinfo{journal}{Phys. Rev.} \textbf{\bibinfo{volume}{C71}},
  \bibinfo{pages}{064901} (\bibinfo{year}{2005}), \eprint{hep-ph/0406278}.

\bibitem[{\citenamefont{Xu et~al.}(2008)\citenamefont{Xu, Greiner, and
  Stocker}}]{Xu:2008dv}
\bibinfo{author}{\bibfnamefont{Z.}~\bibnamefont{Xu}},
  \bibinfo{author}{\bibfnamefont{C.}~\bibnamefont{Greiner}}, \bibnamefont{and}
  \bibinfo{author}{\bibfnamefont{H.}~\bibnamefont{Stocker}},
  \bibinfo{journal}{J. Phys.} \textbf{\bibinfo{volume}{G35}},
  \bibinfo{pages}{104016} (\bibinfo{year}{2008}), \eprint{0807.2986}.

\bibitem[{\citenamefont{Broniowski et~al.}(2008)\citenamefont{Broniowski,
  Chojnacki, Florkowski, and Kisiel}}]{Broniowski:2008vp}
\bibinfo{author}{\bibfnamefont{W.}~\bibnamefont{Broniowski}},
  \bibinfo{author}{\bibfnamefont{M.}~\bibnamefont{Chojnacki}},
  \bibinfo{author}{\bibfnamefont{W.}~\bibnamefont{Florkowski}},
  \bibnamefont{and} \bibinfo{author}{\bibfnamefont{A.}~\bibnamefont{Kisiel}},
  \bibinfo{journal}{Phys. Rev. Lett.} \textbf{\bibinfo{volume}{101}},
  \bibinfo{pages}{022301} (\bibinfo{year}{2008}), \eprint{0801.4361}.

\bibitem[{\citenamefont{McLerran and
  Venugopalan}(1994{\natexlab{a}})}]{McLerran:1993ni}
\bibinfo{author}{\bibfnamefont{L.~D.} \bibnamefont{McLerran}} \bibnamefont{and}
  \bibinfo{author}{\bibfnamefont{R.}~\bibnamefont{Venugopalan}},
  \bibinfo{journal}{Phys. Rev.} \textbf{\bibinfo{volume}{D49}},
  \bibinfo{pages}{2233} (\bibinfo{year}{1994}{\natexlab{a}}),
  \eprint{hep-ph/9309289}.

\bibitem[{\citenamefont{McLerran and
  Venugopalan}(1994{\natexlab{b}})}]{McLerran:1993ka}
\bibinfo{author}{\bibfnamefont{L.~D.} \bibnamefont{McLerran}} \bibnamefont{and}
  \bibinfo{author}{\bibfnamefont{R.}~\bibnamefont{Venugopalan}},
  \bibinfo{journal}{Phys. Rev.} \textbf{\bibinfo{volume}{D49}},
  \bibinfo{pages}{3352} (\bibinfo{year}{1994}{\natexlab{b}}),
  \eprint{hep-ph/9311205}.

\bibitem[{\citenamefont{Kharzeev et~al.}(2005)\citenamefont{Kharzeev, Levin,
  and Nardi}}]{Kharzeev:2001yq}
\bibinfo{author}{\bibfnamefont{D.}~\bibnamefont{Kharzeev}},
  \bibinfo{author}{\bibfnamefont{E.}~\bibnamefont{Levin}}, \bibnamefont{and}
  \bibinfo{author}{\bibfnamefont{M.}~\bibnamefont{Nardi}},
  \bibinfo{journal}{Phys. Rev.} \textbf{\bibinfo{volume}{C71}},
  \bibinfo{pages}{054903} (\bibinfo{year}{2005}), \eprint{hep-ph/0111315}.

\bibitem[{\citenamefont{Pratt}(2008)}]{Pratt:2008qv}
\bibinfo{author}{\bibfnamefont{S.}~\bibnamefont{Pratt}} (\bibinfo{year}{2008}),
  \eprint{0811.3363}.

\bibitem[{\citenamefont{Kolb et~al.}(2000)\citenamefont{Kolb, Sollfrank, and
  Heinz}}]{Kolb:2000sd}
\bibinfo{author}{\bibfnamefont{P.~F.} \bibnamefont{Kolb}},
  \bibinfo{author}{\bibfnamefont{J.}~\bibnamefont{Sollfrank}},
  \bibnamefont{and} \bibinfo{author}{\bibfnamefont{U.~W.} \bibnamefont{Heinz}},
  \bibinfo{journal}{Phys. Rev.} \textbf{\bibinfo{volume}{C62}},
  \bibinfo{pages}{054909} (\bibinfo{year}{2000}), \eprint{hep-ph/0006129}.

\bibitem[{\citenamefont{Jas and Mrowczynski}(2007)}]{Jas:2007rw}
\bibinfo{author}{\bibfnamefont{W.}~\bibnamefont{Jas}} \bibnamefont{and}
  \bibinfo{author}{\bibfnamefont{S.}~\bibnamefont{Mrowczynski}},
  \bibinfo{journal}{Phys. Rev.} \textbf{\bibinfo{volume}{C76}},
  \bibinfo{pages}{044905} (\bibinfo{year}{2007}), \eprint{0706.2273}.

\bibitem[{\citenamefont{Sinyukov}(2006)}]{Sinyukov:2006dw}
\bibinfo{author}{\bibfnamefont{Y.~M.} \bibnamefont{Sinyukov}},
  \bibinfo{journal}{Acta Phys. Polon.} \textbf{\bibinfo{volume}{B37}},
  \bibinfo{pages}{3343} (\bibinfo{year}{2006}).

\bibitem[{\citenamefont{Gyulassy et~al.}(2007)\citenamefont{Gyulassy, Sinyukov,
  Karpenko, and Nazarenko}}]{Gyulassy:2007zz}
\bibinfo{author}{\bibfnamefont{M.}~\bibnamefont{Gyulassy}},
  \bibinfo{author}{\bibfnamefont{Y.~M.} \bibnamefont{Sinyukov}},
  \bibinfo{author}{\bibfnamefont{I.}~\bibnamefont{Karpenko}}, \bibnamefont{and}
  \bibinfo{author}{\bibfnamefont{A.~V.} \bibnamefont{Nazarenko}},
  \bibinfo{journal}{Braz. J. Phys.} \textbf{\bibinfo{volume}{37}},
  \bibinfo{pages}{1031} (\bibinfo{year}{2007}).

\bibitem[{\citenamefont{Sinyukov}(2008)}]{Sinyukov:qm08}
\bibinfo{author}{\bibfnamefont{Y.}~\bibnamefont{Sinyukov}}
  (\bibinfo{year}{2008}), \bibinfo{note}{{talk presented at Quark Matter 2008,
  Jaipur, India, 4-10 February 2008}}.

\bibitem[{\citenamefont{Bialas et~al.}(2008)\citenamefont{Bialas, Chojnacki,
  and Florkowski}}]{Bialas:2007gn}
\bibinfo{author}{\bibfnamefont{A.}~\bibnamefont{Bialas}},
  \bibinfo{author}{\bibfnamefont{M.}~\bibnamefont{Chojnacki}},
  \bibnamefont{and}
  \bibinfo{author}{\bibfnamefont{W.}~\bibnamefont{Florkowski}},
  \bibinfo{journal}{Phys. Lett.} \textbf{\bibinfo{volume}{B661}},
  \bibinfo{pages}{325} (\bibinfo{year}{2008}), \eprint{0708.1076}.

\bibitem[{\citenamefont{Broniowski and Florkowski}(2001)}]{Broniowski:2001we}
\bibinfo{author}{\bibfnamefont{W.}~\bibnamefont{Broniowski}} \bibnamefont{and}
  \bibinfo{author}{\bibfnamefont{W.}~\bibnamefont{Florkowski}},
  \bibinfo{journal}{Phys. Rev. Lett.} \textbf{\bibinfo{volume}{87}},
  \bibinfo{pages}{272302} (\bibinfo{year}{2001}), \eprint{nucl-th/0106050}.

\bibitem[{\citenamefont{Baran et~al.}(2004)\citenamefont{Baran, Broniowski, and
  Florkowski}}]{Baran:2003nm}
\bibinfo{author}{\bibfnamefont{A.}~\bibnamefont{Baran}},
  \bibinfo{author}{\bibfnamefont{W.}~\bibnamefont{Broniowski}},
  \bibnamefont{and}
  \bibinfo{author}{\bibfnamefont{W.}~\bibnamefont{Florkowski}},
  \bibinfo{journal}{Acta Phys. Polon.} \textbf{\bibinfo{volume}{B35}},
  \bibinfo{pages}{779} (\bibinfo{year}{2004}), \eprint{nucl-th/0305075}.

\bibitem[{\citenamefont{Kisiel et~al.}(2006)\citenamefont{Kisiel, Taluc,
  Broniowski, and Florkowski}}]{Kisiel:2005hn}
\bibinfo{author}{\bibfnamefont{A.}~\bibnamefont{Kisiel}},
  \bibinfo{author}{\bibfnamefont{T.}~\bibnamefont{Taluc}},
  \bibinfo{author}{\bibfnamefont{W.}~\bibnamefont{Broniowski}},
  \bibnamefont{and}
  \bibinfo{author}{\bibfnamefont{W.}~\bibnamefont{Florkowski}},
  \bibinfo{journal}{Comput. Phys. Commun.} \textbf{\bibinfo{volume}{174}},
  \bibinfo{pages}{669} (\bibinfo{year}{2006}), \eprint{nucl-th/0504047}.

\bibitem[{\citenamefont{Kisiel et~al.}(2008)\citenamefont{Kisiel, Broniowski,
  Chojnacki, and Florkowski}}]{Kisiel:2008ws}
\bibinfo{author}{\bibfnamefont{A.}~\bibnamefont{Kisiel}},
  \bibinfo{author}{\bibfnamefont{W.}~\bibnamefont{Broniowski}},
  \bibinfo{author}{\bibfnamefont{M.}~\bibnamefont{Chojnacki}},
  \bibnamefont{and}
  \bibinfo{author}{\bibfnamefont{W.}~\bibnamefont{Florkowski}}
  (\bibinfo{year}{2008}), \eprint{0808.3363}.

\bibitem[{\citenamefont{Chojnacki et~al.}(2005)\citenamefont{Chojnacki,
  Florkowski, and Csorgo}}]{Chojnacki:2004ec}
\bibinfo{author}{\bibfnamefont{M.}~\bibnamefont{Chojnacki}},
  \bibinfo{author}{\bibfnamefont{W.}~\bibnamefont{Florkowski}},
  \bibnamefont{and} \bibinfo{author}{\bibfnamefont{T.}~\bibnamefont{Csorgo}},
  \bibinfo{journal}{Phys. Rev.} \textbf{\bibinfo{volume}{C71}},
  \bibinfo{pages}{044902} (\bibinfo{year}{2005}), \eprint{nucl-th/0410036}.

\bibitem[{\citenamefont{Lisa and Pratt}(2008)}]{Lisa:2008gf}
\bibinfo{author}{\bibfnamefont{M.~A.} \bibnamefont{Lisa}} \bibnamefont{and}
  \bibinfo{author}{\bibfnamefont{S.}~\bibnamefont{Pratt}}
  (\bibinfo{year}{2008}), \eprint{0811.1352}.

\bibitem[{\citenamefont{Broniowski et~al.}(2009)\citenamefont{Broniowski,
  Rybczynski, and Bozek}}]{Broniowski:2007nz}
\bibinfo{author}{\bibfnamefont{W.}~\bibnamefont{Broniowski}},
  \bibinfo{author}{\bibfnamefont{M.}~\bibnamefont{Rybczynski}},
  \bibnamefont{and} \bibinfo{author}{\bibfnamefont{P.}~\bibnamefont{Bozek}},
  \bibinfo{journal}{Comput. Phys. Commun.} \textbf{\bibinfo{volume}{180}},
  \bibinfo{pages}{69} (\bibinfo{year}{2009}), \eprint{0710.5731}.

\bibitem[{\citenamefont{Bozek}(2008)}]{Bozek:2007di}
\bibinfo{author}{\bibfnamefont{P.}~\bibnamefont{Bozek}}, \bibinfo{journal}{Acta
  Phys. Polon.} \textbf{\bibinfo{volume}{B39}}, \bibinfo{pages}{1375}
  (\bibinfo{year}{2008}), \eprint{0711.2889}.

\bibitem[{\citenamefont{Aoki et~al.}(2006)\citenamefont{Aoki, Fodor, Katz, and
  Szabo}}]{Aoki:2005vt}
\bibinfo{author}{\bibfnamefont{Y.}~\bibnamefont{Aoki}},
  \bibinfo{author}{\bibfnamefont{Z.}~\bibnamefont{Fodor}},
  \bibinfo{author}{\bibfnamefont{S.~D.} \bibnamefont{Katz}}, \bibnamefont{and}
  \bibinfo{author}{\bibfnamefont{K.~K.} \bibnamefont{Szabo}},
  \bibinfo{journal}{JHEP} \textbf{\bibinfo{volume}{01}}, \bibinfo{pages}{089}
  (\bibinfo{year}{2006}), \eprint{hep-lat/0510084}.

\bibitem[{\citenamefont{Chojnacki et~al.}(2008)\citenamefont{Chojnacki,
  Florkowski, Broniowski, and Kisiel}}]{Chojnacki:2007rq}
\bibinfo{author}{\bibfnamefont{M.}~\bibnamefont{Chojnacki}},
  \bibinfo{author}{\bibfnamefont{W.}~\bibnamefont{Florkowski}},
  \bibinfo{author}{\bibfnamefont{W.}~\bibnamefont{Broniowski}},
  \bibnamefont{and} \bibinfo{author}{\bibfnamefont{A.}~\bibnamefont{Kisiel}},
  \bibinfo{journal}{Phys. Rev.} \textbf{\bibinfo{volume}{C78}},
  \bibinfo{pages}{014905} (\bibinfo{year}{2008}), \eprint{0712.0947}.

\bibitem[{\citenamefont{Adler et~al.}(2004)}]{Adler:2003cb}
\bibinfo{author}{\bibfnamefont{S.~S.} \bibnamefont{Adler}} \bibnamefont{et~al.}
  (\bibinfo{collaboration}{PHENIX}), \bibinfo{journal}{Phys. Rev.}
  \textbf{\bibinfo{volume}{C69}}, \bibinfo{pages}{034909}
  (\bibinfo{year}{2004}), \eprint{nucl-ex/0307022}.

\bibitem[{\citenamefont{Adler et~al.}(2003)}]{Adler:2003kt}
\bibinfo{author}{\bibfnamefont{S.~S.} \bibnamefont{Adler}} \bibnamefont{et~al.}
  (\bibinfo{collaboration}{PHENIX}), \bibinfo{journal}{Phys. Rev. Lett.}
  \textbf{\bibinfo{volume}{91}}, \bibinfo{pages}{182301}
  (\bibinfo{year}{2003}), \eprint{nucl-ex/0305013}.

\bibitem[{\citenamefont{Voloshin}(1998)}]{Voloshin:azHBT}
\bibinfo{author}{\bibfnamefont{S.~V.} \bibnamefont{Voloshin}},
  \bibinfo{journal}{{LBNL annual report R20}}  (\bibinfo{year}{1998}),
  \bibinfo{note}{{http://ie.lbl.gov/nsd1999/rnc/RNC.htm}}.

\bibitem[{\citenamefont{Lisa et~al.}(2000)}]{Lisa:2000xj}
\bibinfo{author}{\bibfnamefont{M.~A.} \bibnamefont{Lisa}} \bibnamefont{et~al.}
  (\bibinfo{collaboration}{E895}), \bibinfo{journal}{Phys. Lett.}
  \textbf{\bibinfo{volume}{B496}}, \bibinfo{pages}{1} (\bibinfo{year}{2000}),
  \eprint{nucl-ex/0007022}.

\bibitem[{\citenamefont{Adams et~al.}(2005)}]{Adams:2004yc}
\bibinfo{author}{\bibfnamefont{J.}~\bibnamefont{Adams}} \bibnamefont{et~al.}
  (\bibinfo{collaboration}{STAR}), \bibinfo{journal}{Phys. Rev.}
  \textbf{\bibinfo{volume}{C71}}, \bibinfo{pages}{044906}
  (\bibinfo{year}{2005}), \eprint{nucl-ex/0411036}.

\end{thebibliography}

\end{document}